\documentclass[a4paper,11pt]{article}
\usepackage{jheppub} 
\usepackage{lineno}
\usepackage{float}
\usepackage{comment}

\usepackage{graphicx}
\usepackage{feynmp-auto}
\usepackage{tikz-feynman}
\tikzfeynmanset{compat=1.1.0}

\usepackage{amsmath}
\allowdisplaybreaks

\usepackage{booktabs} 

\usepackage{amssymb,amsmath,amsfonts,nccmath,amsthm}
\usepackage{hyperref}
\usepackage{IEEEtrantools}
\usepackage{float}
\usepackage{graphicx}
\usepackage{xcolor}
\usepackage{tensor}
\usepackage[utf8]{inputenc}
\usepackage{subfig}
\usepackage{bm,bbm,bbold}
\usepackage{scalerel}[2016/12/29]
\usepackage{enumitem}
\usepackage{multirow}
\usepackage{tabularx}
\usepackage{booktabs}
\usepackage{array, makecell}
\usepackage{mathtools}
\usepackage{algorithm}
\usepackage[noend]{algpseudocode}
\usepackage{placeins}

\title{\boldmath Testing General Relativity Through Gravitational Wave Classification: A Convolutional Neural Network Framework}

\author[a]{Lavinia Heisenberg,}
\author[a]{Shayan Hemmatyar,}
\author[b]{Hector Villarrubia-Rojo}

\affiliation[a]{Institute for Theoretical Physics, Heidelberg University,
Philosophenweg 16, 69120 Heidelberg, Germany}

\affiliation[b]{Departamento de F\'isica Te\'orica and IPARCOS,
Facultad de Ciencias F\'isicas, Universidad Complutense de Madrid,
Plaza de las Ciencias 1, 28040 Madrid, Spain}

\emailAdd{heisenberg@thphys.uni-heidelberg.de}
\emailAdd{hemmatyar@thphys.uni-heidelberg.de}
\emailAdd{hectorvi@ucm.es}

\abstract{
We present a machine learning framework for testing general relativity (GR) with gravitational wave signals from binary black hole mergers. Using the source parameters of 173 BBH events from the GWTC catalog as a realistic astrophysical population, we generate simulated GR waveforms and construct beyond GR (BGR) waveforms by applying controlled phase deformations. We introduce a response function formalism that provides a systematic framework for quantifying how any observable responds to modifications of GR. We train convolutional neural networks (CNNs) on two input representations: whitened waveforms and a response function type observable derived from the waveform mismatch, which isolates the effect of phase deviations from the bulk signal. Using response functions as the CNN input improves the classification sensitivity by a factor of approximately 33 compared to whitened waveforms, demonstrating that the choice of observable representation is as important as the classifier architecture. We study the fundamental limits of this classification through Bayes optimal error analysis, averaging methods that reveal coherent patterns hidden in noise, and a comparison between CNN accuracy and a single feature classifier as a proxy for human performance. At all deformation scales, the CNN outperforms the best single feature approach. We extend the framework to physically motivated theories using the parameterized post Einsteinian (ppE) formalism and apply it to massive gravity, where the classifier detects deviations for graviton masses of order $m_g \sim 10^{-23}\;\mathrm{eV}/c^2$ with aLIGO design sensitivity.
}

\keywords{gravitational waves, tests of general relativity, modified gravity, convolutional neural networks, massive graviton, response functions}

\begin{document}
\maketitle
\flushbottom

\newtheorem{definition}{Definition}[section]
\newtheorem{example}{Example}[section]
\newtheorem{theorem}{Theorem}[section]
\newtheorem{corollary}{Corollary}[section]
\newtheorem{lemma}{Lemma}[section]
\newtheorem{remark}{Remark}[section]
\newtheorem{proposition}{Proposition}[section]


\section{Introduction}
\label{sec:introduction}

The detection of gravitational waves (GWs) from binary black hole mergers by LIGO and Virgo~\cite{abbott_observation_2016,abbott_gwtc1_2019} has opened a direct way to test general relativity (GR) in the strong field regime. The phase and amplitude of the signal emitted during the inspiral carry information about the dynamics of spacetime, and comparing these observations with theoretical predictions allows us to look for deviations that could point to modifications of gravity~\cite{abbott_tests_2021,LIGOScientific:2026qni,LIGOScientific:2026fcf,yunes_fundamentaltheoreticalbiasgravitationalwaveastrophysicsparameterizedposteinsteinianframework_2009,Heisenberg:2018vsk}.

Standard methods for testing GR with gravitational wave data use Bayesian parameter estimation~\cite{abbott_testsgr_gw150914_2016,abbott_tests_2021,LIGOScientific:2026qni,LIGOScientific:2026fcf}. These methods compare observed signals with template waveforms that contain additional parameters beyond GR and then constrain the size of possible deviations. The parameterized post Einsteinian (ppE) framework~\cite{yunes_fundamentaltheoreticalbiasgravitationalwaveastrophysicsparameterizedposteinsteinianframework_2009,tahura_parameterizedposteinsteiniangravitationalwaveformsvariousmodifiedtheoriesgravity_2018,Maggio:2022hre,Pompili:2025cdc,Piarulli:2025rvr} provides a way to encode such deviations in the waveform phase in a theory independent manner. Machine learning offers a complementary approach to these established methods~\cite{cuoco_enhancing_2021,huerta_accelerated_2021,xie_neuralposteinsteinianframeworkefficienttheoryagnostictestsgeneralrelativitygravitationalwaves_2024}. A key advantage of neural networks is their flexibility in accepting different input representations. Standard Bayesian parameter estimation operates on the detector strain and compares it with waveform templates through a likelihood function whose form is tied to the noise model. A convolutional neural network (CNN), by contrast, learns the mapping from any input to a classification directly from training data, without requiring an analytic likelihood. This means that observables other than the raw waveform, such as residuals, response functions, or any other derived quantity, can be directly fed into the network without reformulating the underlying statistical framework. The neural network serves as a flexible platform for exploring which representations of the data are most sensitive to deviations from GR.

In this work, we develop a classification framework using convolutional neural networks (CNNs) to distinguish GR from BGR gravitational wave signals. We construct BGR waveforms by modifying the phase of the dominant $(\ell, m) = (2,2)$ mode with a Gaussian deformation, controlled by a parameter $\beta$ that sets the deviation strength. This toy model allows us to systematically study the sensitivity of the classifier as a function of the deformation size without committing to a specific theory. We consider two input representations for the CNN: whitened waveforms and response functions. Inspired by the response function approach introduced in~\cite{heisenberg_simultaneouslysolving08tensionslatedarkenergy_2023,heisenberg_canlatetimeextensionssolve08tensions_2022} in the context of cosmological tensions, we develop a response function formalism for gravitational waveforms that provides a general framework for quantifying how any observable responds to phase deformations in a frequency dependent way. We apply this formalism to the waveform mismatch, which yields a frequency domain representation that isolates the effect of the phase deviation from the bulk signal. We show that using response functions as input leads to a large improvement in classification accuracy.

We also study the fundamental limits of this classification. The Bayes optimal error~\cite{devroye_probabilistic_1996,bishop_pattern_2006} sets a lower bound on the classification error due to the overlap between the GR and BGR distributions that no classifier can overcome. We compare the CNN accuracy with an optimal single feature classifier that uses only one hand picked number from the response function, namely the mean amplitude near the deformation frequency, and classifies based on an optimal threshold on that value. Our results show that the CNN consistently outperforms this single feature baseline at every deformation scale, and that both converge to $50\%$ (random guessing) as $\beta \to 0$. Additionally, we show that averaging response functions across many events reveals coherent BGR patterns that cancel in GR, although this averaging requires prior knowledge of the labels and cannot serve as a standalone classification method.

Finally, we extend the analysis from the Gaussian toy model to physically motivated modified theories of gravity. Using the ppE framework~\cite{yunes_fundamentaltheoreticalbiasgravitationalwaveastrophysicsparameterizedposteinsteinianframework_2009}, we show how various theories, including massive gravity~\cite{rham_massivegravity_2014,derham_resummation_2011,will_gravitonmass_1998}, can be cast into the same phase deformation framework, and we train the CNN pipeline on massive gravity to evaluate its detectability.

The paper is organized as follows. In Section~\ref{sec:waveforms}, we review the frequency domain gravitational waveform and describe how we construct BGR waveforms through phase deformations. Section~\ref{sec:data_methodology} describes the dataset, noise model, and our event level split methodology. In Section~\ref{sec:waveform_classification}, we present classification results using whitened waveforms. Section~\ref{sec:response_functions} introduces the response function formalism and compares its classification performance with the waveform approach. The limits of classification are discussed in Section~\ref{sec:limits}, including visual diagnostics, averaging methods, Bayes optimal error, and the CNN versus single feature classifier comparison. In Section~\ref{sec:ppe}, we apply the framework to massive gravity using the ppE formalism. We conclude in Section~\ref{sec:conclusion}. Details of the neural network architecture and evaluation metrics are provided in Appendices~\ref{app:nn_architecture} and~\ref{app:eval_metrics}.


\section{Gravitational Waveforms and Beyond GR Extensions}
\label{sec:waveforms}

In this section we review the frequency domain representation of gravitational waveforms from compact binary mergers, introduce a general framework for parameterizing deviations from GR, and define the Gaussian toy model that we use throughout this work.

\subsection{Frequency Domain Waveforms}
\label{sec:freq_domain_waveforms}

The gravitational wave strain $h(t)$ emitted by a compact binary is decomposed into two polarization modes~\cite{maggiore_gravitational_2008}:
\begin{equation}
    h(t) \equiv h_{+}(t) - i\, h_{\times}(t)\,.
\end{equation}
In the time domain, the strain can be expanded in terms of spin weighted spherical harmonics~\cite{mehta_accurate_2017,khan_frequencydomaingravitationalwavesnonprecessingblackholebinariesiiphenomenologicalmodeladvanceddetectorera_2016}:
\begin{equation}
    h\left(t;\, \boldsymbol{\lambda},\, \iota,\, \varphi_c\right) = \sum_{\ell=2}^{+\infty} \sum_{m=-\ell}^{\ell}\,_{-2}Y_{\ell m}\left(\iota, \varphi_c\right)\, h_{\ell m}(t, \boldsymbol{\lambda})\,,
\end{equation}
where $\boldsymbol{\lambda}$ denotes the intrinsic parameters of the binary (component masses $m_1$, $m_2$ and spins $\chi_1$, $\chi_2$), $\iota$ is the inclination angle, and $\varphi_c$ is the coalescence phase. In the frequency domain, the two polarizations can be written as sums over the individual modes $\tilde{h}_{\ell m}(f)$. Following~\cite{mehta_accurate_2017} (see their Appendix~C for the full derivation), the result is
\begin{equation}
    \tilde{h}_{+}(f) = \sum_{\ell=2}^{+\infty} \sum_{m=1}^{\ell} \left[(-1)^{\ell}\, \frac{d_{2}^{\ell,-m}(\iota)}{d_{2}^{\ell m}(\iota)} + 1\right]\, Y_{-2}^{\ell m}(\iota, \varphi_c)\, \tilde{h}_{\ell m}(f)\,,
    \label{eq:hplus_modes}
\end{equation}
\begin{equation}
    \tilde{h}_{\times}(f) = -i \sum_{\ell=2}^{+\infty} \sum_{m=1}^{\ell} \left[(-1)^{\ell}\, \frac{d_{2}^{\ell,-m}(\iota)}{d_{2}^{\ell m}(\iota)} - 1\right]\, Y_{-2}^{\ell m}(\iota, \varphi_c)\, \tilde{h}_{\ell m}(f)\,,
    \label{eq:hcross_modes}
\end{equation}
where $d_{2}^{\ell m}(\iota)$ are the Wigner $d$-matrix elements evaluated at the inclination angle. The angular factors encode the geometry of the source relative to the detector, and all of the dynamical content is carried by the frequency domain modes $\tilde{h}_{\ell m}(f)$. Each mode can be written as~\cite{mehta_accurate_2017}
\begin{equation}
    \tilde{h}_{\ell m}(f) = A_{\ell m}(f)\, e^{i \psi_{\ell m}(f)}\,,
    \label{eq:freq_mode}
\end{equation}
where $A_{\ell m}(f)$ and $\psi_{\ell m}(f)$ are the frequency domain amplitude and phase. The phase $\psi_{\ell m}(f)$ is the main object of our study, as it carries detailed information about the orbital dynamics and is sensitive to modifications of the underlying theory of gravity.

\subsection{Parameterized Phase and Amplitude Deviations}
\label{sec:param_deviations}

We now introduce a general framework for parameterizing deviations from GR in the gravitational waveform. The phase and amplitude of each mode are functions of frequency and the intrinsic binary parameters $\boldsymbol{\lambda} = \{\eta,\, M,\, \chi_{\mathrm{S}},\, \chi_{\mathrm{A}}\}$, where
\begin{equation}
    \eta = \frac{m_1 m_2}{(m_1 + m_2)^2}\,, \qquad M = m_1 + m_2\,, \qquad
    \chi_{\mathrm{S}} = \frac{\chi_1 + \chi_2}{2}\,, \qquad \chi_{\mathrm{A}} = \frac{\chi_1 - \chi_2}{2}\,.
    \label{eq:binary_params}
\end{equation}
We allow the phase and amplitude to differ from their GR predictions by small deviations $\delta \psi_{\ell m}$ and $\delta A_{\ell m}$~\cite{pan_inspiral-merger-ringdown_2011,broeck_phenomenologyamplitudecorrectedpostnewtoniangravitationalwaveformscompactbinaryinspiralsignaltonoiseratios_2007}:
\begin{equation}
    \psi_{\ell m}(f;\, \boldsymbol{\lambda}) = \psi_{\ell m}^{(\mathrm{GR})}(f;\, \boldsymbol{\lambda}) + \delta \psi_{\ell m}(f)\,,
    \label{eq:phaseshift_1}
\end{equation}
\begin{equation}
    A_{\ell m}(f;\, \boldsymbol{\lambda}) = A_{\ell m}^{(\mathrm{GR})}(f;\, \boldsymbol{\lambda}) + \delta A_{\ell m}(f)\,.
\end{equation}
The deviations $\delta \psi_{\ell m}(f)$ and $\delta A_{\ell m}(f)$ are in general functions of frequency. In some modified theories they also depend on $\boldsymbol{\lambda}$ (for example, through the chirp mass in the ppE framework), but in this work we first study purely frequency dependent deformations before specializing to specific theories in Section~\ref{sec:ppe}.

\label{sec:gaussian_toy_model}
To construct beyond GR (BGR) waveforms in a controlled way, we modify only the dominant $(\ell, m) = (2,2)$ mode of the GR waveform. This mode carries most of the signal power for the quasi circular, non precessing binaries we consider. Higher order modes can be important for asymmetric mass ratios or large inclination angles, but restricting the modification to the $(2,2)$ mode is sufficient for our purposes and keeps the framework simple. The full GR waveform in the frequency domain is a sum over all modes:
\begin{equation}
    h_{\mathrm{GR}}(f) = \sum_{\ell, m} \tilde{h}_{\ell m, \mathrm{GR}}(f)\,,
\end{equation}
where each $\tilde{h}_{\ell m, \mathrm{GR}}(f) = A_{\ell m}(f)\, e^{i\psi_{\ell m}(f)}$ as defined in Eq.~\eqref{eq:freq_mode}. To construct a BGR waveform, we add a phase shift $\beta\, \delta\psi_{22}(f)$ to the $(2,2)$ mode only, where $\beta$ is a free parameter controlling the strength of the deviation:
\begin{equation}
    h_{\mathrm{BGR}}(f) = h_{\mathrm{GR}}(f) + \tilde{h}_{22, \mathrm{GR}}(f)\left(e^{i\,\beta\,\delta\psi_{22}(f)} - 1\right).
    \label{eq:bgr_exact}
\end{equation}
This expression is exact: it replaces the original $(2,2)$ mode with a phase shifted version while leaving all other modes unchanged. For small $\beta\,\delta\psi_{22}$, expanding the exponential to first order gives
\begin{equation}
    e^{i\,\beta\,\delta\psi_{22}(f)} - 1 \approx i\,\beta\,\delta\psi_{22}(f)\,,
\end{equation}
so that
\begin{equation}
    h_{\mathrm{BGR}}(f) \approx h_{\mathrm{GR}}(f) + i\,\beta\,\tilde{h}_{22, \mathrm{GR}}(f)\,\delta\psi_{22}(f)\,.
    \label{eq:general_modification}
\end{equation}
This linearized form is useful for analytic calculations, such as the derivation of the response function in Section~\ref{sec:response_functions}. In our numerical implementation, we use the exact expression Eq.~\eqref{eq:bgr_exact}.

As a controlled toy model for testing and validating the classification framework, we choose a Gaussian deformation centered in the inspiral regime:
\begin{equation}
    \delta\psi_{22}(f) = \exp\left[-\frac{(f - 50)^2}{100}\right],
    \label{eq:gaussian_shift}
\end{equation}
which is peaked at $50$ Hz with a width of $10$ Hz. This form is not meant to represent any specific modified gravity theory. It is a deliberately simple, localized deformation used only to benchmark the pipeline across a wide range of deformation strengths, before we turn to physically motivated theories via the ppE framework in Section~\ref{sec:ppe}. The specific values of the center and width are not physically motivated; they serve only to localize the deformation in a frequency range where the detector is sensitive and the inspiral dynamics dominate.

\begin{figure}[htbp]
    \centering
    \includegraphics[width=0.8\textwidth]{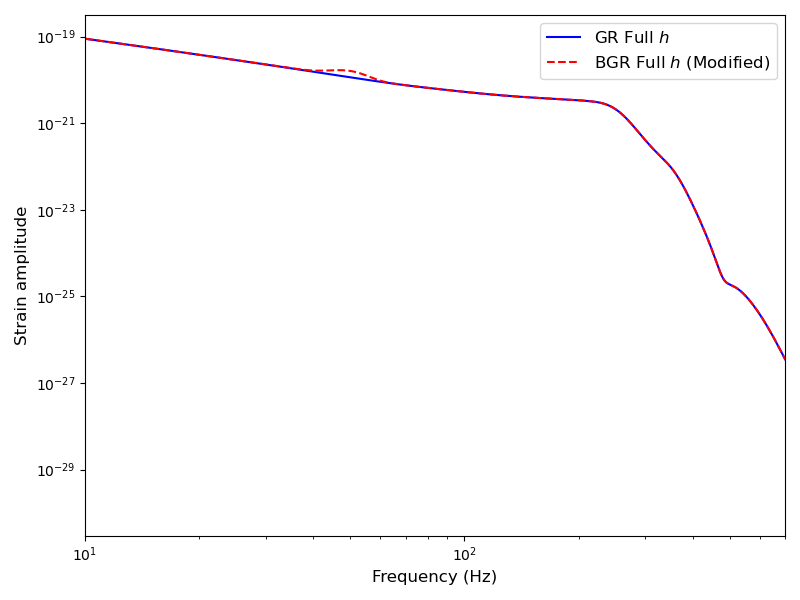}
    \caption{Illustration of the phase modification applied to the $(2,2)$ mode. The red dashed line shows the BGR waveform with a small phase shift in the inspiral (low frequency) regime compared to the GR waveform (blue).}
    \label{fig:Deviation}
\end{figure}

The BGR waveform is then obtained by inserting Eq.~\eqref{eq:gaussian_shift} into Eq.~\eqref{eq:bgr_exact}. The parameter $\beta$ controls how large the deviation is: for $\beta = 1$ the modification is strong and easily detectable, while for $\beta \ll 1$ the GR and BGR waveforms become increasingly similar and harder to distinguish. By varying $\beta$, we can systematically probe the sensitivity of our classification framework across different deformation scales.

The Gaussian form is chosen for simplicity and because it naturally vanishes at high frequencies, which is a requirement in our framework since we only study modifications in the inspiral regime. The specific shape of $\delta\psi_{22}(f)$ does not correspond to any particular modified gravity theory; it serves as a first controlled test case for the classification pipeline. In Section~\ref{sec:ppe}, we replace this toy deformation with the phase corrections predicted by specific theories using the ppE framework.


\section{Data, Noise, and Methodology}
\label{sec:data_methodology}

The parameterized waveform framework developed in Section~\ref{sec:waveforms} provides a way to construct BGR signals for any choice of binary parameters. To train a classifier that generalizes to real observations, we need to apply this framework to a population of events that reflects the actual distribution of binary black hole mergers observed by current detectors. In this section, we describe how we build such a dataset. We start from the source parameters (masses, spins, distances) of the 173 BBH events catalogued in GWTC-1, GWTC-2, and GWTC-3~\cite{abbott_gwtc1_2019,abbott_gwtc2_2021,abbott_gwtc3_2023}, which define a realistic astrophysical population. Throughout this work, all waveforms are simulated injections generated from these catalog parameters; no real detector strain data is analyzed. We first augment the parameter set by perturbing the source parameters to broaden the coverage of the binary parameter space. For each augmented parameter set, we then generate a simulated GR waveform and construct the corresponding BGR waveform using the phase deformation of Eq.~\eqref{eq:bgr_exact}. Finally, we inject colored Gaussian noise drawn from the Advanced LIGO design sensitivity curve to simulate realistic observational conditions. Finally, we define a train/validation/test split at the event level to prevent data leakage between augmented copies of the same event.

\subsection{BBH Events and Data Augmentation}
\label{sec:bbh_events}

For each of the 173 catalogued parameter sets, we extract the source frame component masses $m_1$, $m_2$, the effective spin parameter $\chi_{\mathrm{eff}}$, the redshift $z$, and the luminosity distance $d_L$. Waveforms are generated with the PyCBC package~\cite{alexnitz_gwastropycbcv233releasepycbc_2024} using the IMRPhenomXHM approximant~\cite{garcia-quiros_multimodefrequencydomainmodelgravitationalwavesignalnonprecessingblackholebinaries_2020}, with a low frequency cutoff of $f_{\mathrm{low}} = 10$ Hz and a frequency resolution of $\Delta f = 1$ Hz. The luminosity distance is passed to the waveform generator so that the signal amplitude reflects realistic observational conditions.

With only 173 events, the network would see a limited set of parameter combinations during training, which increases the risk of overfitting to the specific waveform morphologies present in the catalog rather than learning general features that distinguish GR from BGR signals. We apply data augmentation by generating 10 perturbed copies of each event with small Gaussian shifts in the component masses and effective spin. This augmentation serves two purposes. First, it acts as a regularizer: by presenting the network with slightly different versions of the same underlying event, we force it to learn features that are stable under small changes in the binary parameters rather than memorizing exact spectral shapes. This is analogous to standard image augmentation techniques (small rotations, crops, color jitter), where the perturbations are also small but improve generalization by breaking exact symmetries in the training data~\cite{shorten_surveyimagedataaugmentation_2019}. Second, the augmented events remain anchored to the observed population of binary black hole mergers, so the classifier is trained on the region of parameter space that is astrophysically relevant for current detectors rather than on an arbitrary prior over binary parameters. Although the perturbations are small (3--5\%), each augmented event produces a waveform with a different chirp mass, merger frequency, and overall spectral shape. Specifically, for each augmented copy we draw a perturbation fraction $p$ uniformly between 3\% and 5\%, and then sample
\begin{equation}
    m_1' = m_1 + \mathcal{N}(0,\, p\, m_1)\,, \qquad m_2' = m_2 + \mathcal{N}(0,\, p\, m_2)\,,
\end{equation}
\begin{equation}
    \chi_{\mathrm{eff}}' = \chi_{\mathrm{eff}} + \mathcal{N}(0,\, p\, |\chi_{\mathrm{eff}}| + 0.01)\,,
\end{equation}
where $\mathcal{N}(0, \sigma)$ denotes a Gaussian with zero mean and standard deviation $\sigma$. We enforce $m_1' \geq m_2'$ and $|\chi_{\mathrm{eff}}'| < 1$ after perturbation, ensuring that the component masses remain ordered by convention and that the effective spin stays within the physical range set by the Kerr bound on individual black hole spins. Including the original events, this gives 11 variants per event and expands the dataset to $173 \times 11 = 1903$ events in total.

\begin{figure}[h!]
    \centering
    \includegraphics[width=0.85\textwidth]{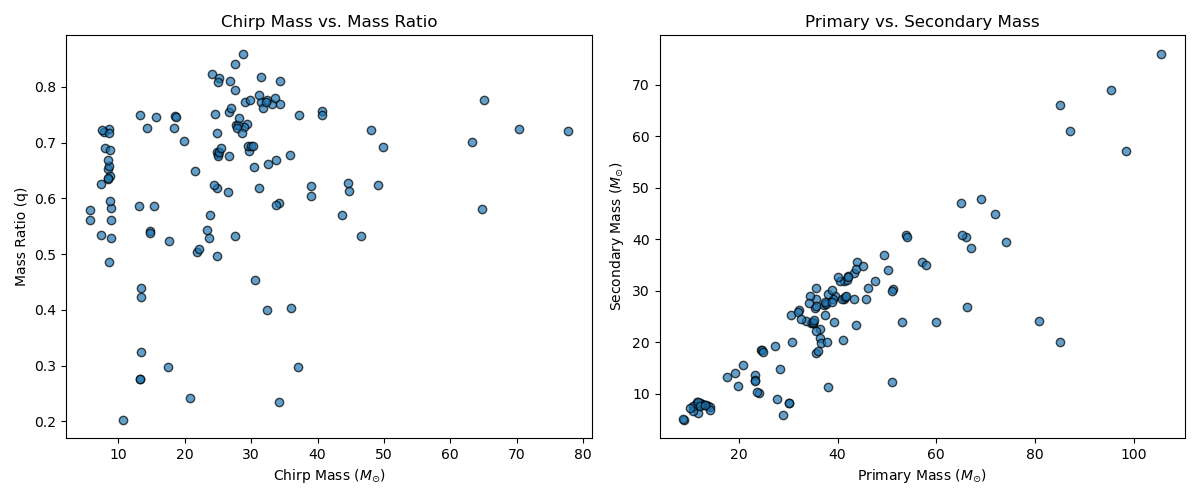}\\[6pt]
    \includegraphics[width=0.85\textwidth]{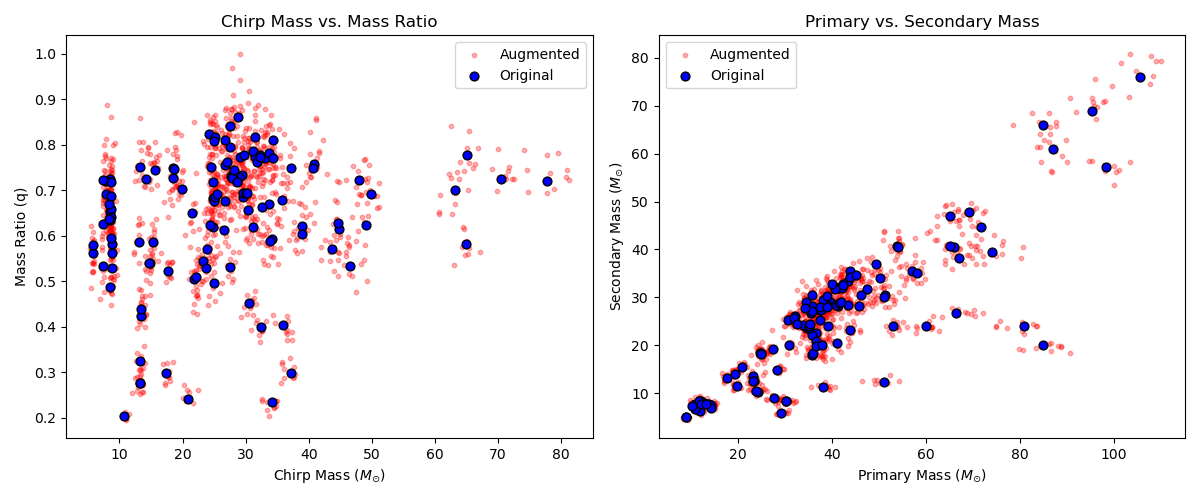}
    \caption{Distribution of BBH source parameters before (top) and after (bottom) data augmentation. Each row shows the chirp mass versus mass ratio (left) and primary versus secondary mass (right). The top row contains the 173 catalogued parameter sets. The bottom row shows the same sets (blue) together with their 10 augmented copies each (red). The augmented parameter sets cluster around the originals, broadening the local parameter space coverage.}
    \label{fig:augmentation}
\end{figure}

Figure~\ref{fig:augmentation} illustrates the effect of the augmentation. The top panel of each subplot shows the 173 original catalog events, while the bottom panel shows the same events together with their augmented copies. The augmented points form local clouds around each original event, broadening the coverage of the parameter space while remaining anchored to the observed population.

\subsection{Detector Noise and Signal to Noise Ratio}
\label{sec:detector_noise}

To simulate realistic observational conditions, we add colored Gaussian noise to the waveforms. The noise in gravitational wave detectors is characterized by the power spectral density (PSD), which describes how the noise power is distributed across frequencies~\cite{moore_gravitationalwavesensitivitycurves_2015}. The noise is ``colored'' because the PSD is not flat: the detector is much more sensitive at some frequencies than others, so the noise power varies strongly across the band. For a noise realization $n(t)$ with Fourier transform $\tilde{n}(f)$, the PSD is defined as
\begin{equation}
    S_n(f) = \lim_{T \to \infty} \frac{1}{T} \left\langle |\tilde{n}(f)|^2 \right\rangle\,,
\end{equation}
where $\langle \cdot \rangle$ denotes an ensemble average. We model the noise as stationary (its statistical properties do not change with time) and Gaussian (each frequency component is drawn from a Gaussian distribution). Under these assumptions, different frequency components are uncorrelated:
\begin{equation}
    \left\langle \tilde{n}(f)\, \tilde{n}^*(f') \right\rangle = \delta(f - f')\, S_n(f)\,.
\end{equation}
Real detector noise also contains non stationary and non Gaussian features such as glitches and instrumental transients, which we do not model in this work.

We use the Advanced LIGO design sensitivity curve~\cite{aasi_advancedligo_2015}. For each event, we generate 20 independent noise realizations by drawing frequency domain noise from a Gaussian distribution with variance determined by $S_n(f)$. The noisy waveform is then
\begin{equation}
    \tilde{h}_{\mathrm{noisy}}(f) = \tilde{h}_{\mathrm{clean}}(f) + \tilde{n}(f)\,.
\end{equation}
This noise is added independently to both the GR and BGR waveforms, with different random seeds for each. Before passing the waveforms to the neural network, we whiten them by dividing by $\sqrt{S_n(f)}$, which normalizes the noise to be approximately uniform across frequencies.

\begin{figure}[htbp]
    \centering
    \includegraphics[width=0.85\textwidth]{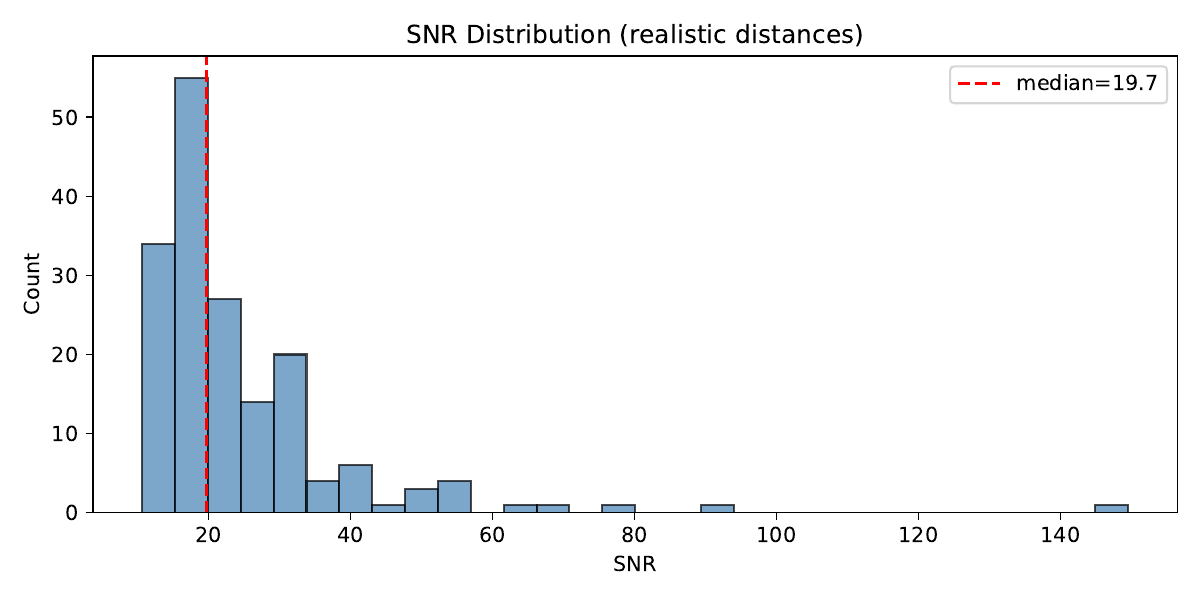}
    \caption{Distribution of optimal SNR values for the simulated injections used in this work. The SNR is computed using the Advanced LIGO design sensitivity curve with the waveforms placed at their catalog luminosity distances.}
    \label{fig:snr_distribution}
\end{figure}

The signal to noise ratio (SNR) of a gravitational wave signal quantifies its strength relative to the detector noise. It is defined using the noise weighted inner product~\cite{moore_gravitationalwavesensitivitycurves_2015}:
\begin{equation}
    (A \mid B) = 4\, \mathrm{Re} \int_0^\infty \frac{A^*(f)\, B(f)}{S_n(f)}\, df\,,
    \label{eq:inner_product}
\end{equation}
where $S_n(f)$ is the PSD. The optimal SNR of a signal $h(f)$ is then
\begin{equation}
    \mathrm{SNR} = \sqrt{(h \mid h)} = \sqrt{4 \int_0^\infty \frac{|h(f)|^2}{S_n(f)}\, df}\,.
    \label{eq:snr}
\end{equation}
Because we use realistic luminosity distances, the SNR of each simulated injection reflects the detectability of a source with the corresponding parameters. The median SNR across the 173 parameter sets is approximately $19.7$, and the distribution is shown in Figure~\ref{fig:snr_distribution}. Injections at larger distances have lower SNR, making the classification task harder for those signals.

\subsection{Event Level Train/Validation/Test Split}
\label{sec:event_split}

We split the data at the \emph{event level}: the 173 original BBH events are randomly divided into 121 training events, 25 validation events, and 27 test events (approximately 70\%/15\%/15\%). All augmented copies of a given event are assigned to the same set as the original. This means that no event, and no augmented version of that event, appears in more than one set.

This event level split is important because augmented copies of the same event share very similar waveform morphology. If augmented copies were allowed to appear in both the training and test sets, the classifier could achieve high accuracy simply by memorizing the waveform shapes of individual events, rather than learning general features that distinguish GR from BGR signals. This type of data leakage leads to inflated accuracy estimates that do not generalize to new, unseen events.

After augmentation and noise injection, the final dataset sizes are:
\begin{itemize}
    \item \textbf{Training}: $121 \times 11 \times 20 \times 2 = 53{,}240$ samples,
    \item \textbf{Validation}: $25 \times 11 \times 20 \times 2 = 11{,}000$ samples,
    \item \textbf{Test}: $27 \times 11 \times 20 \times 2 = 11{,}880$ samples,
\end{itemize}
where the factor of 11 comes from the original event plus 10 augmented copies, 20 is the number of noise realizations, and the factor of 2 accounts for the GR and BGR classes. Each waveform is whitened by dividing by $\sqrt{S_n(f)}$ and represented by the real and imaginary parts of the resulting strain. The waveforms are generated at $\Delta f = 1$~Hz resolution, giving 391 frequency bins from 10 to 400~Hz. These are resampled onto a uniform grid of 500 points over the same range to provide a fixed input dimension for the CNN. Each channel is independently normalized to zero mean and unit variance (z score normalization), producing a two channel input of shape $(500, 2)$.

To ensure robustness of our results, we train the classifier with 4 different random seeds and report the mean and standard deviation of the accuracy across these runs.


\section{Waveform Classification}
\label{sec:waveform_classification}
\label{sec:cnn_architecture}

We classify GR and BGR waveforms using a one dimensional convolutional neural network (CNN). The input to the network is the two channel whitened waveform described in Section~\ref{sec:event_split}, with shape $(500, 2)$.

\begin{figure}[htbp]
    \centering
    \includegraphics[width=\textwidth]{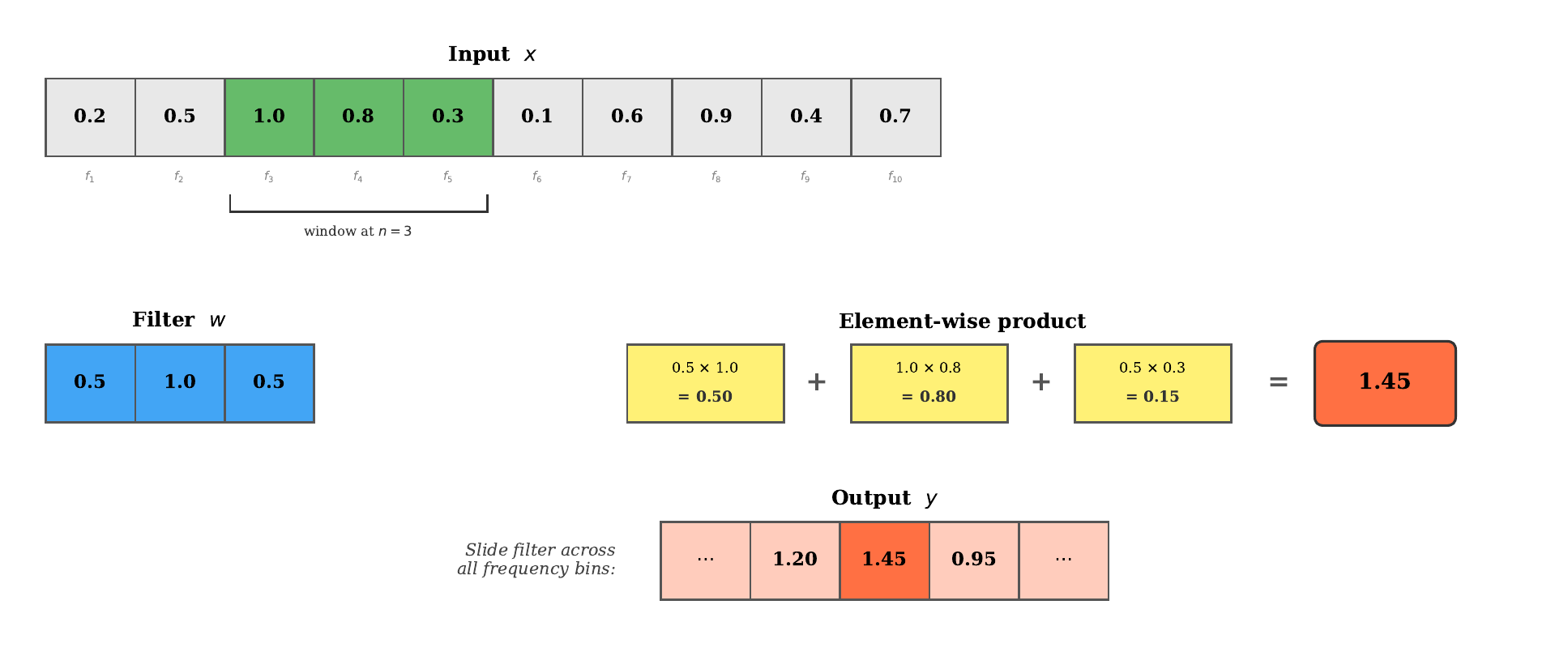}
    \caption{Illustration of a 1D convolution. A filter of kernel size 3 slides across the input frequency series. At each position, the filter weights are multiplied element-wise with the input values in the window (highlighted in green), and the products are summed to produce one entry in the output feature map. The same filter weights are reused at every position, which is the source of translation invariance and parameter efficiency.}
    \label{fig:conv1d_schematic}
\end{figure}

The choice of a CNN over a standard fully connected (dense) network is motivated by the structure of the input data. Both whitened waveforms and response functions are frequency series in which physically meaningful features, such as a coherent bump from a phase deformation or the chirp structure of the signal, are localized in groups of neighboring frequency bins. A CNN exploits this locality by applying small convolutional filters that slide across the frequency axis and detect local patterns regardless of where they appear in the band (Figure~\ref{fig:conv1d_schematic}). This translation invariance is a natural match for our problem, since we do not know in advance at which frequency a deviation from GR will be most visible.

CNNs also require far fewer trainable parameters than a fully connected architecture of comparable capacity. To illustrate, a single dense layer connecting all 1000 input values ($500 \times 2$ channels) to 128 hidden units would already require $128{,}000$ weights. By contrast, our first convolutional layer uses 32 filters of kernel size 11 on 2 input channels, giving only $32 \times (11 \times 2 + 1) = 736$ parameters, because the same filter weights are shared across all frequency bins. Our entire CNN has approximately 67,500 total parameters, which is less than what a single fully connected layer would need. This parameter efficiency reduces the risk of overfitting and makes the network well suited for scaling to more complex tasks, such as multiclass classification among several modified gravity theories, where the convolutional backbone can be reused and only the output layer needs to change.

For each value of the deformation strength $\beta$, we train the network 4 times with different random seeds and report the mean and standard deviation of the test accuracy. Throughout this work, we adopt a 95\% accuracy threshold as the criterion for reliable classification. This is a conventional choice: below this level, the classifier makes errors on more than one in twenty samples, which we consider too frequent for a confident detection of beyond GR physics. The full architecture details, including layer specifications, regularization, and training hyperparameters, are given in Appendix~\ref{app:nn_architecture}.

\subsection{Mismatch Between GR and BGR Waveforms}
\label{sec:mismatch}

Before presenting the classification results, we introduce the match and mismatch between two waveforms, which provide a standard measure of waveform similarity in gravitational wave data analysis~\cite{moore_gravitationalwavesensitivitycurves_2015}. Using the inner product defined in Eq.~\eqref{eq:inner_product}, the match between two waveforms $A(f)$ and $B(f)$ is
\begin{equation}
    \mathcal{M}(A, B) = \max_{t_0, \phi_0} \frac{(A \mid B)}{\sqrt{(A \mid A)(B \mid B)}}\,,
    \label{eq:match}
\end{equation}
where the maximization is over relative time and phase shifts. The mismatch is then defined as
\begin{equation}
    1 - \mathcal{M} = 1 - \max_{t_0, \phi_0} \frac{(A \mid B)}{\sqrt{(A \mid A)(B \mid B)}}\,.
    \label{eq:mismatch}
\end{equation}
A mismatch close to zero means the two waveforms are nearly identical, while a larger mismatch indicates a stronger deviation. For each value of $\beta$, we compute the mismatch between the GR waveform and its BGR counterpart for every event in the dataset and then average over all events.

\subsection{Waveform Classification Results}
\label{sec:waveform_results}

We train the CNN for each value of $\beta$ and evaluate its performance on the held out test set. Table~\ref{tab:wf_results} summarizes the classification accuracy, F1 score, and average mismatch for each $\beta$. The results are averaged over 4 random seeds.

\begin{table}[htbp]
\centering
\caption{Waveform classification results for different values of $\beta$. The accuracy and F1 score are reported as mean $\pm$ standard deviation over 4 random seeds. The average mismatch between GR and BGR waveforms is also shown. The mismatch is not strictly monotonic at large $\beta$ because the Gaussian phase deformation is localized in frequency and the mismatch saturates (see text); small variations in the saturated regime reflect the event-averaged oscillatory behavior of the phase factor $e^{i\beta\,\delta\psi(f)}$.}
\label{tab:wf_results}
\begin{tabular}{cccc}
\toprule
$\beta$ & Accuracy (\%) & F1 Score & Avg.\ Mismatch \\
\midrule
20   & $96.1 \pm 0.2$ & $0.961 \pm 0.002$ & $1.09 \times 10^{-1}$ \\
15   & $94.3 \pm 0.8$ & $0.941 \pm 0.009$ & $1.11 \times 10^{-1}$ \\
10   & $93.3 \pm 1.0$ & $0.930 \pm 0.011$ & $1.15 \times 10^{-1}$ \\
5    & $93.2 \pm 0.3$ & $0.929 \pm 0.003$ & $9.87 \times 10^{-2}$ \\
1    & $80.3 \pm 0.3$ & $0.793 \pm 0.022$ & $1.40 \times 10^{-2}$ \\
0.7  & $76.2 \pm 0.2$ & $0.712 \pm 0.002$ & $7.10 \times 10^{-3}$ \\
0.5  & $70.3 \pm 0.5$ & $0.624 \pm 0.014$ & $3.78 \times 10^{-3}$ \\
0.3  & $50.1 \pm 0.2$ & $\sim 0.50$ & $1.38 \times 10^{-3}$ \\
0.1  & $50.0 \pm 0.0$ & $\sim 0.50$ & $1.54 \times 10^{-4}$ \\
0.01 & $50.0 \pm 0.0$ & $\sim 0.50$ & $1.54 \times 10^{-6}$ \\
\bottomrule
\end{tabular}
\end{table}

The 95\% accuracy threshold is crossed at approximately $\beta \approx 20$. At $\beta = 20$ the classifier reaches $96.1\%$, while at $\beta = 15$ the accuracy is $94.3\%$, just below the threshold. Accuracies at $\beta = 5$ and $\beta = 10$ are similar at around 93\%, indicating a plateau where moderately large deformations are distinguishable from GR but not reliably enough to meet the 95\% criterion. The plateau exists because at large $\beta$ the phase factor $e^{i\beta\,\delta\psi_{22}(f)}$ oscillates rapidly across frequency bins. After whitening and noise injection, these rapid oscillations become difficult to distinguish from noise fluctuations, limiting the achievable accuracy even for very strong deformations.

Below $\beta = 1$, the classification performance degrades rapidly. At $\beta = 1$ the accuracy is $80.3\%$, still well above random chance, but by $\beta = 0.3$ the accuracy has dropped to $50.1\%$, which is indistinguishable from random guessing. For all $\beta \leq 0.3$, the network completely fails to separate GR from BGR waveforms, and the F1 score converges to $\sim 0.50$, consistent with random chance.

\begin{figure}[htbp]
    \centering
    \includegraphics[width=\textwidth]{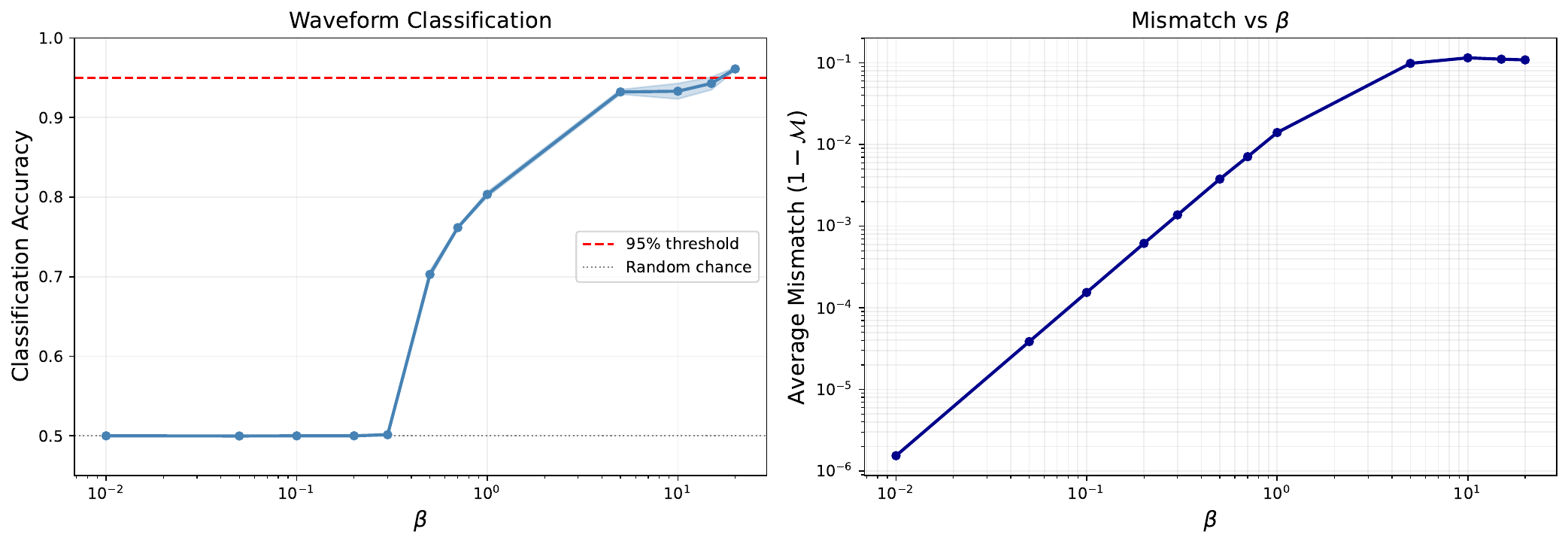}
    \caption{Left: classification accuracy as a function of $\beta$ for waveform classification. The dashed red line marks the 95\% accuracy threshold, the dotted gray line indicates random chance (50\%), and the shaded band shows the $\pm 1\sigma$ spread over 4 random seeds. The threshold is crossed at approximately $\beta \approx 20$. Right: average mismatch between GR and BGR waveforms as a function of $\beta$, showing a monotonic increase that saturates around $10^{-1}$.}
    \label{fig:wf_summary}
\end{figure}

Figure~\ref{fig:wf_summary} shows the classification accuracy and average mismatch as a function of $\beta$. The left panel shows a steep transition between $\beta = 1$ and $\beta = 0.3$, where the network goes from partial classification ability to random chance. The shaded band indicates the $\pm 1\sigma$ spread across the 4 random seeds. The right panel shows the average mismatch between GR and BGR waveforms, which grows monotonically with $\beta$ over several orders of magnitude, from $\sim 10^{-6}$ at $\beta = 0.01$ to $\sim 10^{-1}$ at $\beta \gtrsim 5$, where it saturates. This saturation occurs because the Gaussian deformation is localized in frequency: once $\beta$ is large enough to fully scramble the phase within the Gaussian window, the match contribution from those frequencies is effectively zero, but the signal at frequencies outside the deformation region remains unmodified and continues to match the GR template. The saturation level therefore reflects the fraction of the total SNR weighted power that lies outside the Gaussian bump.

\begin{figure}[htbp]
    \centering
    \includegraphics[width=\textwidth]{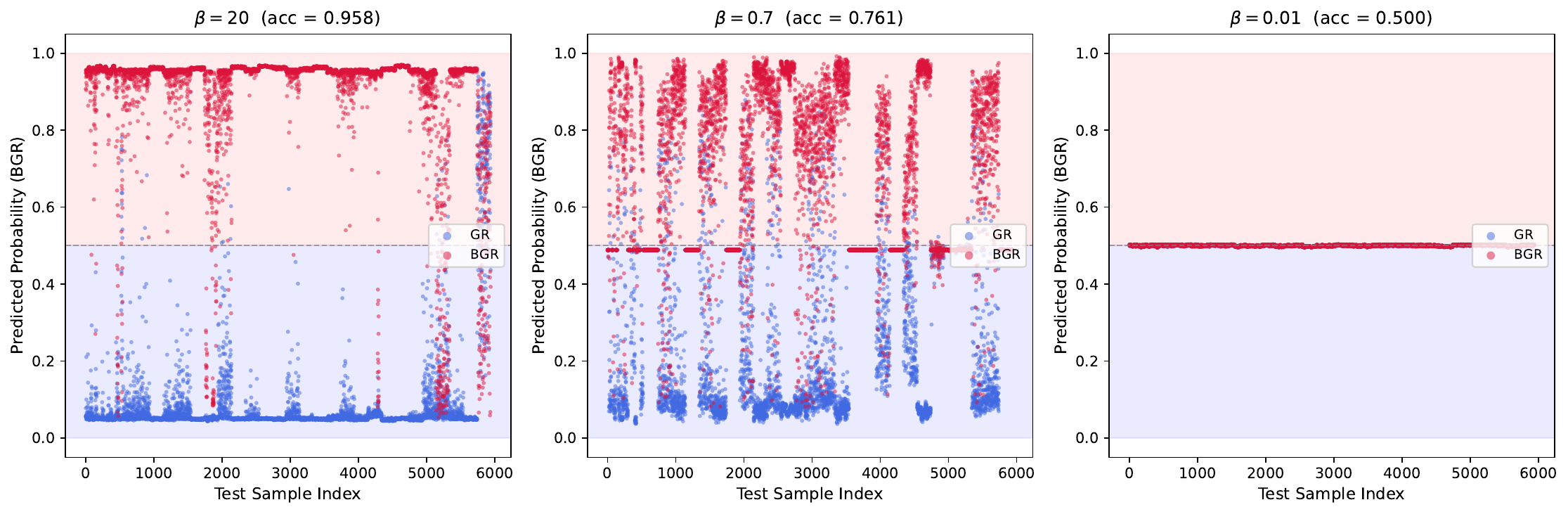}
    \caption{Predicted BGR probability for each test sample at $\beta = 20$ (left), $\beta = 0.7$ (center), and $\beta = 0.01$ (right). Blue and red points correspond to true GR and BGR samples respectively. The dashed line marks the 0.5 decision boundary. As $\beta$ decreases, the two classes become increasingly mixed.}
    \label{fig:wf_scatter}
\end{figure}

To illustrate how the classifier behaves at different deformation scales, Figure~\ref{fig:wf_scatter} shows the predicted BGR probability for each test sample at three representative values of $\beta$. At $\beta = 20$, the GR and BGR samples are clearly separated by the decision boundary at 0.5, with only a few misclassified samples. At $\beta = 0.7$, the predicted probabilities spread across the full range, and the classifier makes frequent errors on both classes. At $\beta = 0.01$, all predictions cluster around 0.5, confirming that the network finds no distinguishing features.

\begin{figure}[htbp]
    \centering
    \includegraphics[width=\textwidth]{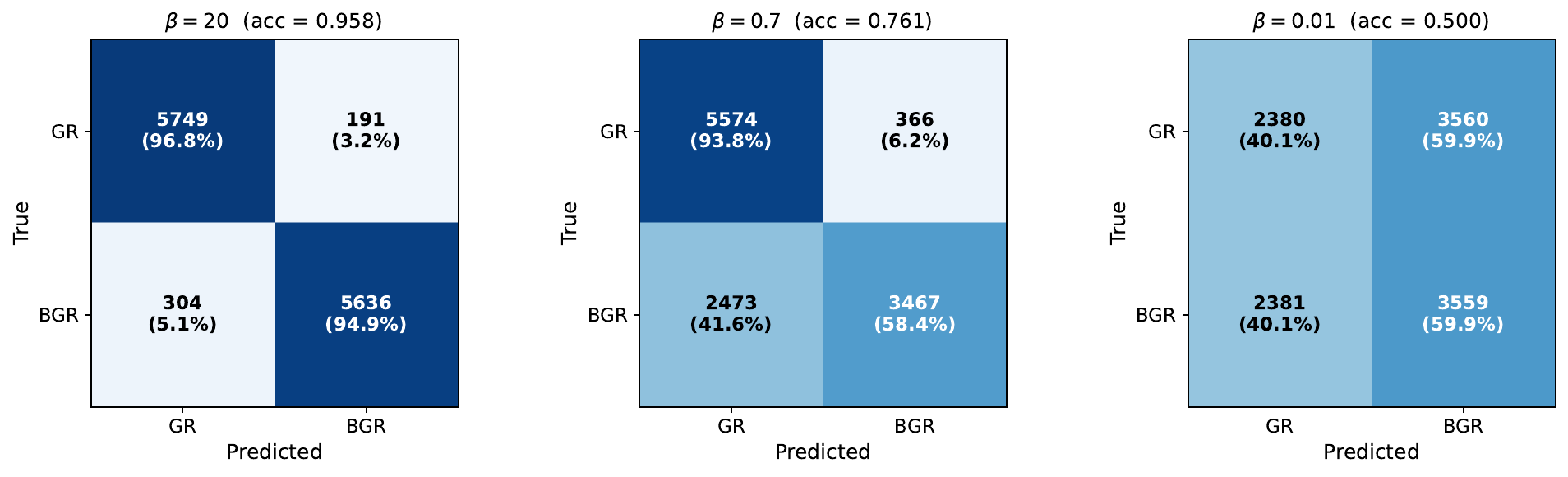}
    \caption{Confusion matrices at $\beta = 20$ (left), $\beta = 0.7$ (center), and $\beta = 0.01$ (right). Percentages indicate the fraction of each true class classified into each predicted category (rows sum to 100\%).}
    \label{fig:wf_confusion}
\end{figure}

Figure~\ref{fig:wf_confusion} shows the confusion matrices for the same three values of $\beta$. At $\beta = 20$, misclassification rates are low (3.2\% false positives and 5.1\% false negatives). At $\beta = 0.7$, the network still correctly identifies most GR samples but misclassifies a large fraction (41.6\%) of BGR samples as GR. At $\beta = 0.01$, the two rows of the confusion matrix are identical: the network produces the same distribution of predictions regardless of the true label, confirming that it has no discriminative ability. However, the predictions are not uniformly split between the two classes: approximately 60\% of all samples are predicted as BGR and 40\% as GR. This class bias arises because the network cannot learn any useful features at this deformation scale, so the sigmoid output drifts slightly above the 0.5 decision threshold during training, causing a systematic preference for one class. The overall accuracy remains exactly 50\% because the bias helps and hurts each class equally.

These results establish a baseline for the sensitivity of waveform classification with the Gaussian toy model. The 95\% accuracy threshold is crossed at approximately $\beta \approx 20$, corresponding to an average mismatch of order $10^{-1}$, meaning that the GR and BGR waveforms must differ at the $\sim 10\%$ level in their noise weighted overlap before the classifier can reliably distinguish them. This is a relatively high threshold, and it motivates the search for a more sensitive observable. In the next section, we introduce the response function formalism, which provides a representation that significantly improves the classifier's sensitivity to smaller deformations.


\section{Response Function Formalism}
\label{sec:response_functions}

The waveform classification results in Section~\ref{sec:waveform_results} show that the CNN requires deformations of order $\beta \approx 20$ to exceed the 95\% accuracy threshold. This motivates the search for a more informative representation of the data. This section serves two purposes. First, we develop the response function formalism for gravitational waveforms, adapting the approach of~\cite{heisenberg_simultaneouslysolving08tensionslatedarkenergy_2023,heisenberg_canlatetimeextensionssolve08tensions_2022} from the cosmological setting to the context of compact binary signals. The formalism provides a general framework for studying how phase deformations propagate into shifts of any observable quantity. Second, we apply this formalism to the waveform mismatch, use the resulting response function as a new input representation for the CNN, and show that it leads to a large improvement in classification sensitivity compared to the whitened waveform approach of Section~\ref{sec:waveform_classification}.

\subsection{General Response Function Formalism}
\label{sec:response_theory}

The presence of a phase deviation $\delta \psi_{\ell m}$ also induces shifts in the estimated binary parameters. When fitting a GR template to a BGR signal, the best fit values of the parameters will be shifted by amounts $\delta \eta$, $\delta M$, $\delta \chi_{\mathrm{S}}$, and $\delta \chi_{\mathrm{A}}$. The total observed shift in phase is then
\begin{equation}
    \Delta \psi_{\ell m} = \delta \psi_{\ell m} +
    \frac{\partial \psi_{\ell m}^{(\mathrm{GR})}}{\partial \eta}\, \delta \eta +
    \frac{\partial \psi_{\ell m}^{(\mathrm{GR})}}{\partial M}\, \delta M +
    \frac{\partial \psi_{\ell m}^{(\mathrm{GR})}}{\partial \chi_{\mathrm{S}}}\, \delta \chi_{\mathrm{S}} +
    \frac{\partial \psi_{\ell m}^{(\mathrm{GR})}}{\partial \chi_{\mathrm{A}}}\, \delta \chi_{\mathrm{A}}\,,
\end{equation}
and similarly for the amplitude. Normalizing the parameter shifts by their values, this becomes
\begin{equation}
\begin{aligned}
    \Delta \psi_{\ell m} = \delta \psi_{\ell m} \;&+\;
    \left(\eta \frac{\partial \psi_{\ell m}^{(\mathrm{GR})}}{\partial \eta}\right) \frac{\delta \eta}{\eta} \;+\;
    \left(M \frac{\partial \psi_{\ell m}^{(\mathrm{GR})}}{\partial M}\right) \frac{\delta M}{M} \\
    &+\;
    \left(\chi_{\mathrm{S}} \frac{\partial \psi_{\ell m}^{(\mathrm{GR})}}{\partial \chi_{\mathrm{S}}}\right) \frac{\delta \chi_{\mathrm{S}}}{\chi_{\mathrm{S}}} \;+\;
    \left(\chi_{\mathrm{A}} \frac{\partial \psi_{\ell m}^{(\mathrm{GR})}}{\partial \chi_{\mathrm{A}}}\right) \frac{\delta \chi_{\mathrm{A}}}{\chi_{\mathrm{A}}}\,,
\end{aligned}
\label{eq:relative_shift}
\end{equation}
with an analogous expression for $\Delta A_{\ell m}$. This decomposition into a direct deformation and parameter induced shifts is the starting point of the response function formalism. The structure of Eq.~\eqref{eq:relative_shift} motivates the introduction of frequency dependent response functions that characterize how phase and amplitude deformations propagate into parameter shifts. We define the following integral relations:
\begin{equation}
    \frac{\delta \lambda_i}{\lambda_i} = \int df\, R_{\lambda_i}(f)\, \delta\psi_{\ell m}(f) + \int df\, \tilde{R}_{\lambda_i}(f)\, \delta A_{\ell m}(f)\,,
    \label{eq:param_response}
\end{equation}
for each parameter $\lambda_i \in \{\eta, M, \chi_{\mathrm{S}}, \chi_{\mathrm{A}}\}$, where $R_{\lambda_i}(f)$ and $\tilde{R}_{\lambda_i}(f)$ are the phase and amplitude response functions respectively. These functions encode how perturbations in the waveform structure map onto shifts in the physical parameters of the binary system.

Once the parameter response functions are known, the formalism can be extended to study how any observable quantity $O(f;\, \boldsymbol{\lambda})$ that depends on the waveform and the binary parameters responds to deformations. The total shift in $O$ can be written as
\begin{equation}
    \Delta O(f) = \sum_i I_{\lambda_i}(f)\, \frac{\delta \lambda_i}{\lambda_i} + \int df'\, R_O(f', f)\, \delta\psi_{\ell m}(f') + \int df'\, \tilde{R}_O(f', f)\, \delta A_{\ell m}(f')\,,
    \label{eq:obs_shift}
\end{equation}
where $I_{\lambda_i}(f) = \lambda_i\, \partial O / \partial \lambda_i$ are weight functions that describe the direct sensitivity of $O$ to parameter shifts, and $R_O(f', f)$, $\tilde{R}_O(f', f)$ are the intrinsic response kernels of the observable to phase and amplitude deformations. As a concrete example, if the observable is the phase itself, $O = \psi_{\ell m}^{(\mathrm{GR})}$, then $I_{\lambda_i}(f) = \lambda_i\, \partial \psi_{\ell m}^{(\mathrm{GR})} / \partial \lambda_i \equiv g_{\lambda_i}(f)$, and Eq.~\eqref{eq:obs_shift} reduces to the total phase shift of Eq.~\eqref{eq:relative_shift}. Substituting the parameter response functions from Eq.~\eqref{eq:param_response}, the total shift takes a compact convolutional form:
\begin{equation}
    \Delta O(f) = \int df'\, \mathcal{R}_O(f', f)\, \delta\psi_{\ell m}(f') + \int df'\, \tilde{\mathcal{R}}_O(f', f)\, \delta A_{\ell m}(f')\,,
    \label{eq:obs_convolution}
\end{equation}
where the effective response kernels are
\begin{equation}
    \mathcal{R}_O(f', f) = \sum_i I_{\lambda_i}(f)\, R_{\lambda_i}(f') + R_O(f', f)\,, \qquad \tilde{\mathcal{R}}_O(f', f) = \sum_i I_{\lambda_i}(f)\, \tilde{R}_{\lambda_i}(f') + \tilde{R}_O(f', f)\,.
    \label{eq:effective_kernels}
\end{equation}
Equation~\eqref{eq:obs_convolution} shows that the shift in any observable is a linear functional of the waveform deformations, with the effective response kernel encoding both the direct effect of the deformation and the indirect effect through parameter shifts. This generality is one of the key strengths of the formalism: by choosing different observables $O$, one can construct different representations of the same underlying deviation, each potentially offering different sensitivity to beyond GR physics.

In this work, we focus on phase deformations confined to the inspiral regime, so that $\delta\psi_{\ell m}(f)$ vanishes at high frequencies. We impose the boundary condition that the total phase shift vanishes at $f_{22}^{\mathrm{peak}}$, the frequency at which the amplitude of the $(2,2)$ mode reaches its maximum, marking the transition from inspiral to merger~\cite{mehta_testsgeneralrelativitygravitationalwaveobservationsusingflexibletheoryindependentmethod_2023}:
\begin{equation}
    \Delta\psi_{\ell m}\!\left(f_{22}^{\mathrm{peak}}\right) = 0\,.
    \label{eq:boundary_condition}
\end{equation}

To see the consequences, first consider a single parameter $\lambda_i$. From Eq.~\eqref{eq:relative_shift}, the total phase shift at $f_{22}^{\mathrm{peak}}$ is
\begin{equation}
    \Delta\psi_{\ell m}\!\left(f_{22}^{\mathrm{peak}}\right) = \delta\psi_{\ell m}\!\left(f_{22}^{\mathrm{peak}}\right) + g_{\lambda_i}\!\left(f_{22}^{\mathrm{peak}}\right) \frac{\delta\lambda_i}{\lambda_i}\,,
\end{equation}
where $g_{\lambda_i}(f) = \lambda_i\, \partial\psi_{\ell m}^{(\mathrm{GR})}/\partial\lambda_i$. Imposing the boundary condition gives
\begin{equation}
    \frac{\delta\lambda_i}{\lambda_i} = -\frac{\delta\psi_{\ell m}\!\left(f_{22}^{\mathrm{peak}}\right)}{g_{\lambda_i}\!\left(f_{22}^{\mathrm{peak}}\right)}\,.
    \label{eq:single_param_bc}
\end{equation}
For deformations that vanish at high frequencies, $\delta\psi_{\ell m}(f_{22}^{\mathrm{peak}}) \approx 0$, and therefore $\delta\lambda_i/\lambda_i \approx 0$.

This argument generalizes to multiple parameters. When all $N$ parameters are present simultaneously, the boundary condition Eq.~\eqref{eq:boundary_condition} provides one constraint, but there are $N$ unknown parameter shifts. However, the parameter shifts are not independent: they are determined by the full frequency dependent deformation through the integral relations Eq.~\eqref{eq:param_response}. If the deformation $\delta\psi_{\ell m}(f)$ has compact support or decays sufficiently fast at high frequencies, so that $\delta\psi_{\ell m}(f) \to 0$ for $f \to f_{22}^{\mathrm{peak}}$, then each integral $\int df\, R_{\lambda_i}(f)\, \delta\psi_{\ell m}(f)$ receives negligible contributions from the high frequency regime. Since the response functions $R_{\lambda_i}(f)$ are smooth and bounded, and the deformation is localized at low frequencies, each parameter shift $\delta\lambda_i/\lambda_i$ is driven entirely by the low frequency content of $\delta\psi_{\ell m}$ and is proportional to the deformation strength $\beta$. In the limit $\beta \to 0$, all parameter shifts vanish simultaneously, regardless of their number.

More concretely, for a Gaussian deformation centered at $f_0$ with width $\sigma$ (Section~\ref{sec:gaussian_toy_model}), the deformation amplitude at $f_{22}^{\mathrm{peak}}$ is suppressed by a factor $\exp\!\left[-(f_{22}^{\mathrm{peak}} - f_0)^2/(2\sigma^2)\right]$, which is exponentially small when $f_{22}^{\mathrm{peak}} \gg f_0 + \sigma$. This holds simultaneously for all parameters.

Under these conditions, the parameter induced terms in Eq.~\eqref{eq:relative_shift} are negligible, and the total phase shift reduces to the direct deformation:
\begin{equation}
    \Delta\psi_{\ell m}(f) \approx \delta\psi_{\ell m}(f)\,.
\end{equation}
Furthermore, since phase accumulates over many gravitational wave cycles, phase deformations are generally more sensitive probes of small deviations than amplitude modifications~\cite{mehta_testsgeneralrelativitygravitationalwaveobservationsusingflexibletheoryindependentmethod_2023}. We therefore restrict to phase only extensions ($\delta A_{\ell m} = 0$) throughout this work, which simplifies the effective response kernel in Eq.~\eqref{eq:effective_kernels} to $\mathcal{R}_O(f', f) \approx R_O(f', f)$.

\subsection{Mismatch Response Function}
\label{sec:mismatch_response_theory}

We now apply the general formalism to a specific observable: the mismatch between a GR template and the observed signal. The mismatch $\bar{\mathcal{M}} = 1 - \mathcal{M}$ quantifies how well the template matches the data (see Eq.~\eqref{eq:match}). This is a natural choice because the mismatch is one of the most widely used figures of merit in gravitational wave data analysis, and its response function can be computed analytically.

Let $\Delta h(f)$ denote a small variation of the model waveform. The first order variation of the mismatch is
\begin{equation}
    \delta\bar{\mathcal{M}} = -(\Delta\hat{h} \mid \hat{s}) + \mathcal{M}\,(\Delta\hat{h} \mid \hat{h})\,,
    \label{eq:mismatch_variation}
\end{equation}
where $\hat{h} = h/\|h\|$ and $\hat{s} = s/\|s\|$ are the normalized waveforms, and $\Delta\hat{h} = \Delta h/\|h\|$ to first order. For the phase only BGR modification of the $(2,2)$ mode, the waveform variation is $\Delta h = i\beta\, \tilde{h}_{22,\mathrm{GR}}\, \delta\psi_{22}$ (from Eq.~\eqref{eq:general_modification}). Substituting into the noise weighted inner product (Eq.~\eqref{eq:inner_product}), the mismatch variation takes the form
\begin{equation}
    \delta\bar{\mathcal{M}} = \beta\, \mathrm{Re}\!\int R(f)\, \delta\psi_{22}(f)\, df\,,
\end{equation}
where $R(f)$ is the mismatch response function:
\begin{equation}
    R(f) = \frac{4i\, h_{22}(f)}{\|h\|\, S_n(f)} \left[-\hat{s}^*(f) + \mathcal{M}\, \hat{h}^*(f)\right].
    \label{eq:response_function}
\end{equation}
Here $h_{22}(f)$ is the $(2,2)$ mode of the GR template, $\|h\| = \sqrt{(h|h)}$ is its norm, $\mathcal{M} = (\hat{h}|\hat{s})$ is the match, and $S_n(f)$ is the detector PSD. This is a specific instance of the general formalism in Eq.~\eqref{eq:obs_convolution}, with the mismatch playing the role of the observable $O$.

The response function $R(f)$ has a key property that makes it useful for classification. When the signal is consistent with GR, meaning $s = h_{\mathrm{GR}} + n$ where $n$ is detector noise, the match $\mathcal{M} \approx 1$ and $\hat{s} \approx \hat{h}$ at high SNR. The bracket in Eq.~\eqref{eq:response_function} then becomes $[-\hat{h}^* + \hat{h}^*] \approx 0$ plus noise contributions, so $R(f)$ is noise-like with no coherent structure. When the signal is a BGR waveform, meaning $s = h_{\mathrm{BGR}} + n$, the match $\mathcal{M} < 1$ and $\hat{s} \neq \hat{h}$, so the bracket is non-zero and $R(f)$ develops structured features that reflect the underlying phase deviation $\delta\psi_{22}$.

This asymmetry between GR (noise-like) and BGR (structured) response functions is the feature that the CNN learns to distinguish. The response function effectively acts as a residual that amplifies the BGR signal relative to the noise background, making smaller deformations detectable.

\subsection{Response Function Classification Results}
\label{sec:mismatch_response}
\label{sec:response_results}

In practice, for each sample in our dataset we compute the response function $R(f)$ using Eq.~\eqref{eq:response_function}, where $h$ is the noiseless GR template and $s$ is the signal (either GR or BGR) with injected detector noise. The real and imaginary parts of $R(f)$ are interpolated onto the same frequency grid of 500 points from 10 to 400~Hz and independently normalized using z score normalization, producing a two channel input of shape $(500, 2)$ for the CNN. The same CNN architecture described in Section~\ref{sec:cnn_architecture} is used.

\begin{figure}[htbp]
    \centering
    \includegraphics[width=\textwidth]{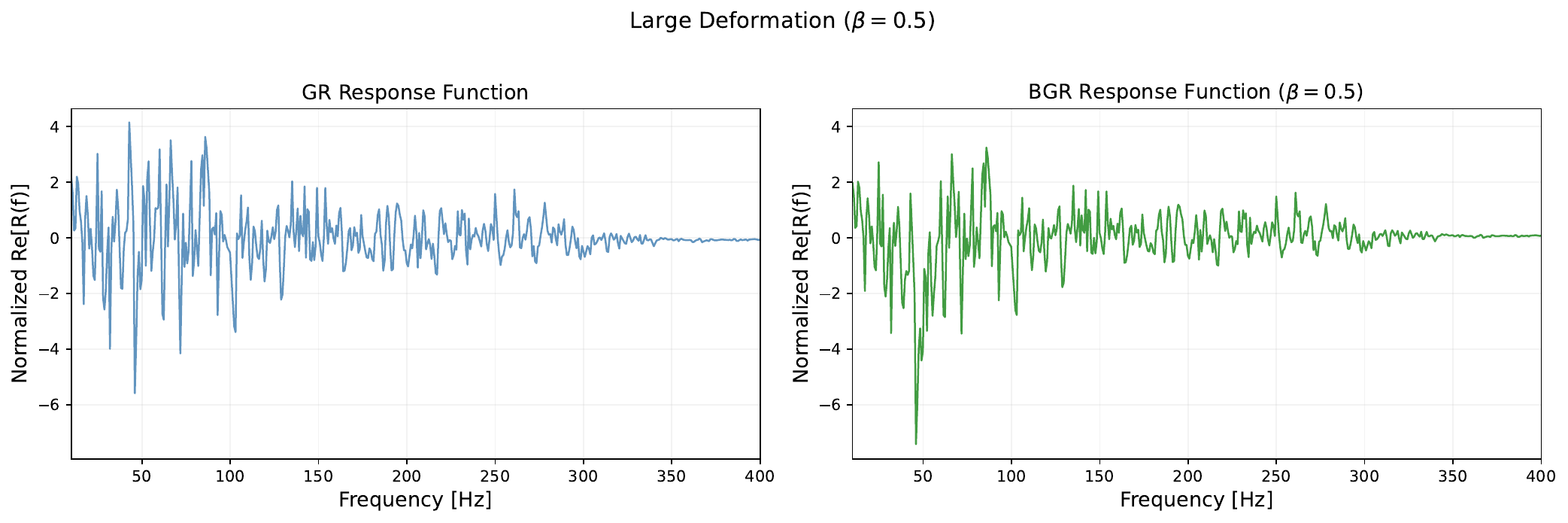}
    \caption{Normalized $\mathrm{Re}[R(f)]$ for a single event at $\beta = 0.5$. Left: GR case, showing noise-like fluctuations with no coherent structure. Right: BGR case, showing the same noise plus a structured component from the phase deviation. The enhanced structure near 50~Hz reflects the Gaussian deformation. The visibility of this structure varies across events; lower SNR events show weaker contrast, which accounts for the $\sim 6\%$ misclassification rate at this $\beta$.}
    \label{fig:response_example}
\end{figure}

Figure~\ref{fig:response_example} illustrates the response function for a single event at $\beta = 0.5$. The GR response function (left) shows stochastic fluctuations with no preferred frequency structure. The BGR response function (right) shows the same noise-like fluctuations but with an additional structured component, visible as enhanced oscillations in the low frequency region around 50~Hz where the Gaussian phase deviation is centered. The visibility of this structure varies across events depending on the SNR; this example shows one of the clearer cases.

We train the same CNN on response function inputs and sweep over $\beta$ values from $1.0$ down to $0.001$. Table~\ref{tab:rf_results} summarizes the classification accuracy, F1 score, and average waveform mismatch for each $\beta$. The mismatch values are the same as in Table~\ref{tab:wf_results}, since they are computed between the original GR and BGR waveforms regardless of the classification input.

\begin{table}[htbp]
\centering
\caption{Response function classification results for different values of $\beta$. The accuracy and F1 score are reported as mean $\pm$ standard deviation over 4 random seeds. The 95\% accuracy threshold is crossed between $\beta = 0.5$ and $\beta = 0.6$.}
\label{tab:rf_results}
\begin{tabular}{cccc}
\toprule
$\beta$ & Accuracy (\%) & F1 Score & Avg.\ Mismatch \\
\midrule
1.0  & $99.3 \pm 0.1$ & $0.993 \pm 0.001$ & $1.40 \times 10^{-2}$ \\
0.8  & $98.4 \pm 0.1$ & $0.984 \pm 0.001$ & $9.17 \times 10^{-3}$ \\
0.7  & $97.5 \pm 0.1$ & $0.975 \pm 0.001$ & $7.10 \times 10^{-3}$ \\
\textbf{0.6}  & $\mathbf{96.1 \pm 0.04}$ & $\mathbf{0.961 \pm 0.000}$ & $\mathbf{5.30 \times 10^{-3}}$ \\
0.5  & $93.9 \pm 0.1$ & $0.938 \pm 0.000$ & $3.78 \times 10^{-3}$ \\
0.4  & $90.4 \pm 0.1$ & $0.902 \pm 0.001$ & $2.43 \times 10^{-3}$ \\
0.3  & $85.6 \pm 0.1$ & $0.852 \pm 0.001$ & $1.38 \times 10^{-3}$ \\
0.2  & $78.2 \pm 0.1$ & $0.775 \pm 0.001$ & $6.16 \times 10^{-4}$ \\
0.1  & $67.4 \pm 0.2$ & $0.660 \pm 0.004$ & $1.54 \times 10^{-4}$ \\
0.05 & $58.9 \pm 0.4$ & $0.585 \pm 0.005$ & $3.86 \times 10^{-5}$ \\
0.01 & $51.0 \pm 0.1$ & $\sim 0.50$ & $1.54 \times 10^{-6}$ \\
0.001 & $50.0 \pm 0.0$ & $\sim 0.50$ & $1.54 \times 10^{-8}$ \\
\bottomrule
\end{tabular}
\end{table}

The response function classifier crosses the 95\% threshold at approximately $\beta \approx 0.6$ (the accuracy is $93.9\%$ at $\beta = 0.5$ and $96.1\%$ at $\beta = 0.6$). This is a dramatic improvement over waveform classification, which required $\beta \approx 20$ to reach the same threshold. The accuracy decreases smoothly with $\beta$, reaching $85.6\%$ at $\beta = 0.3$, $67.4\%$ at $\beta = 0.1$, and approaching random chance for $\beta \leq 0.01$.

\begin{figure}[htbp]
    \centering
    \includegraphics[width=\textwidth]{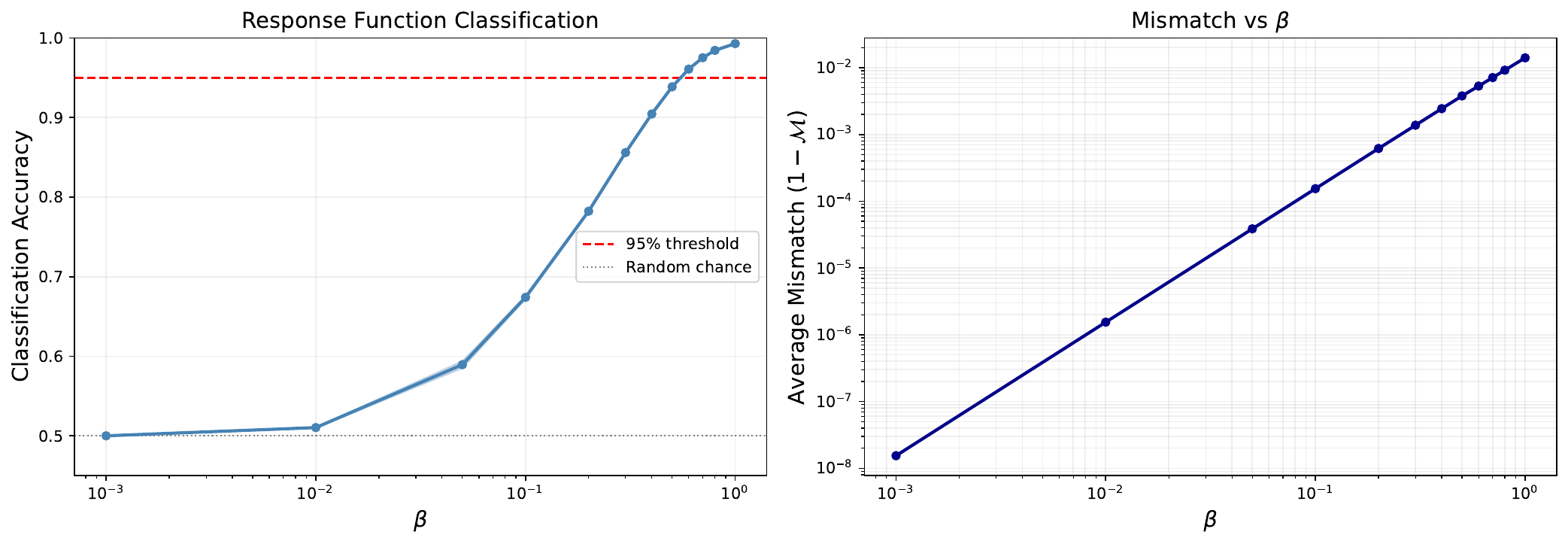}
    \caption{Left: classification accuracy as a function of $\beta$ for response function classification. The 95\% threshold (dashed red) is crossed at approximately $\beta \approx 0.6$. Right: average mismatch between the original GR and BGR waveforms. The response function classifier achieves high accuracy even when the waveform mismatch is very small ($\sim 10^{-3}$).}
    \label{fig:rf_summary}
\end{figure}

Figure~\ref{fig:rf_summary} shows the classification accuracy and the corresponding waveform mismatch as a function of $\beta$.

\begin{figure}[htbp]
    \centering
    \includegraphics[width=\textwidth]{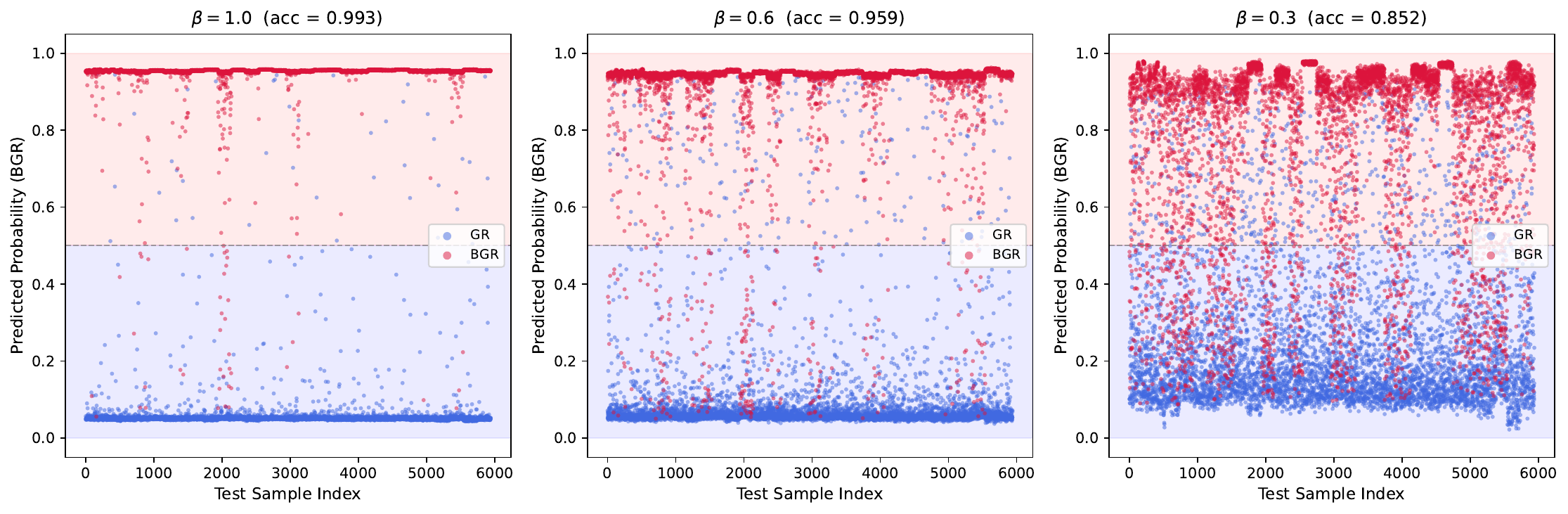}
    \caption{Predicted BGR probability for each test sample in the response function classification at $\beta = 1$ (left), $\beta = 0.6$ (center), and $\beta = 0.3$ (right). Blue and red points correspond to true GR and BGR samples respectively. The dashed line marks the 0.5 decision boundary.}
    \label{fig:rf_scatter}
\end{figure}

Figure~\ref{fig:rf_scatter} shows the predicted BGR probability for test samples at three representative values of $\beta$. At $\beta = 1$, the separation is nearly perfect. At $\beta = 0.6$ (the 95\% threshold), the classes remain well separated with only a few misclassifications. At $\beta = 0.3$, more samples fall near the decision boundary.

\begin{figure}[htbp]
    \centering
    \includegraphics[width=\textwidth]{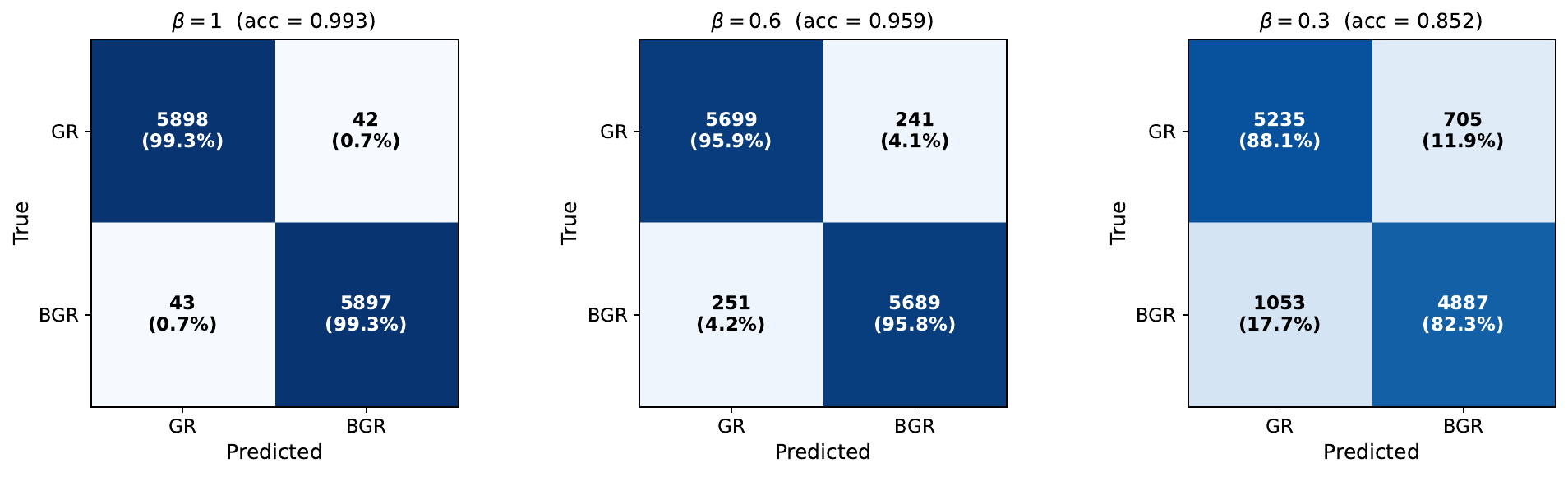}
    \caption{Confusion matrices for the response function classification at $\beta = 1$ (left), $\beta = 0.6$ (center), and $\beta = 0.3$ (right). Percentages indicate the fraction of each true class classified into each predicted category (rows sum to 100\%).}
    \label{fig:rf_confusion}
\end{figure}

Figure~\ref{fig:rf_confusion} shows the confusion matrices for the same three values of $\beta$. At $\beta = 1$, misclassification rates are approximately $0.7\%$ for both classes, consistent with the near perfect scatter plot separation. At $\beta = 0.6$, the false positive and false negative rates are approximately $4\%$ each, nearly symmetric between the two classes. At $\beta = 0.3$, the error rates increase, with $11.9\%$ false positives and $17.7\%$ false negatives, indicating that BGR samples are somewhat harder to classify correctly at this deformation scale. The classifier still performs well above random chance.

\subsection{Comparison of Input Representations}
\label{sec:response_vs_waveforms}

We compare the waveform and response function classifiers at the same values of $\beta$. The response function approach consistently outperforms waveform classification at every deformation scale. The most striking result is the 95\% detection threshold: waveform classification requires $\beta \approx 20$ to reach 95\% accuracy, while response function classification achieves this at approximately $\beta \approx 0.6$, an improvement by a factor of approximately 33. At $\beta = 1$, where waveform classification achieves only $80.3\%$, the response function classifier reaches $99.3\%$. At $\beta = 0.3$, waveform classification has already failed ($50.1\%$), while the response function classifier still achieves $85.6\%$.

This improvement arises because the response function isolates the effect of the phase deviation from the bulk waveform signal. In the waveform representation, the BGR modification is a small perturbation on top of a large GR signal, and the CNN must detect this perturbation in the presence of noise. In the response function representation, the GR component is reduced to noise-like fluctuations, and only the BGR deviation contributes coherent structure. This makes the classification task fundamentally easier.

A natural question is whether this improvement is due to the specific algebraic form of the response function, or whether a simpler construction that also removes the bulk GR component from the observed signal would already capture most of the gain. The simplest such construction is the whitened residual between the observed signal $s(f)$ and the GR template $h_{\mathrm{GR}}(f)$,
\begin{equation}
    r(f) = \frac{s(f) - h_{\mathrm{GR}}(f)}{\sqrt{S_n(f)}}\,.
    \label{eq:residual}
\end{equation}
When the observed signal is consistent with GR, meaning $s = h_{\mathrm{GR}} + n$, the residual reduces to pure whitened noise,
\begin{equation}
    r(f) = \frac{n(f)}{\sqrt{S_n(f)}}\,.
\end{equation}
When it contains a beyond GR modification, meaning $s = h_{\mathrm{BGR}} + n$, the residual contains the whitened BGR deformation plus the same noise,
\begin{equation}
    r(f) = \frac{\delta h(f) + n(f)}{\sqrt{S_n(f)}}\,, \qquad \delta h(f) \equiv h_{\mathrm{BGR}}(f) - h_{\mathrm{GR}}(f)\,.
\end{equation}
The residual shares the noise-like versus structured asymmetry of the response function but does not include the additional $h_{22}(f)/(\|h\|^2 S_n(f))$ weighting. To test whether this weighting matters, we train the same CNN on residual features constructed exactly as above. Table~\ref{tab:comparison} reports the results alongside the waveform and response function classifiers.

\begin{table}[htbp]
\centering
\caption{Comparison of classification accuracy (\%) between the three input representations at the same $\beta$ values. Residual inputs outperform whitened waveforms at every $\beta$, confirming that subtracting the GR template exposes the deformation. The response function outperforms both by a further 20 to 30 percentage points across $\beta$, showing that the $h_{22}(f)/(\|h\|^2 S_n(f))$ weighting in Eq.~\eqref{eq:response_function} contributes a substantial additional improvement on top of the residual.}
\label{tab:comparison}
\begin{tabular}{cccc}
\toprule
$\beta$ & Waveform Acc.\ & Residual Acc.\ & Response Function Acc.\ \\
\midrule
1.0  & $80.3$ & $91.6$ & $99.3$ \\
0.7  & $76.2$ & $79.4$ & $97.5$ \\
0.5  & $70.3$ & $69.4$ & $93.9$ \\
0.3  & $50.1$ & $59.7$ & $85.6$ \\
0.1  & $50.0$ & $50.2$ & $67.4$ \\
0.05 & $50.0$ & $50.0$ & $58.9$ \\
0.01 & $50.0$ & $50.0$ & $51.0$ \\
\bottomrule
\end{tabular}
\end{table}

The residual classifier shows a modest improvement over whitened waveforms at larger $\beta$ (for example 91.6\% vs 80.3\% at $\beta = 1$ and 59.7\% vs 50.1\% at $\beta = 0.3$) and ties with it at intermediate $\beta$. Both fail below $\beta \sim 0.1$. The response function retains a substantial advantage across the entire $\beta$ range, outperforming the residual by 20 to 30 percentage points. This decomposition shows that the gain of the response function framework comes from two separate contributions. Subtracting the GR template removes the bulk signal and exposes the deformation, which is the mechanism underlying Eq.~\eqref{eq:residual}. The extra $h_{22}(f)/(\|h\|^2 S_n(f))$ weighting in the response function then amplifies the deformation in the frequency bins where the signal carries the most information and suppresses it elsewhere, and this second step provides most of the remaining improvement.


\section{Limits of Classification}
\label{sec:limits}

The response function classifier significantly outperforms the waveform approach, but its accuracy still degrades as $\beta$ decreases. In this section, we investigate the fundamental and practical limits of this classification. We begin by visually comparing individual response functions at different deformation scales to understand when the BGR signal becomes invisible to the human eye. We then show that averaging over many samples can recover the coherent BGR pattern even at very small $\beta$, confirming that the signal is present but buried in noise. Next, we relate the observed accuracy to the Bayes optimal error, which sets a fundamental lower bound on the classification error due to the overlap between the GR and BGR distributions. Finally, we compare the CNN to a classifier that uses only one hand picked number from the response function to quantify how much the network gains by using all frequency bins simultaneously.

\subsection{Visual Diagnostics and Coherent Pattern Recovery}
\label{sec:visual_diagnostics}
\label{sec:averaging_methods}

\begin{figure}[htbp]
    \centering
    \includegraphics[width=0.75\textwidth]{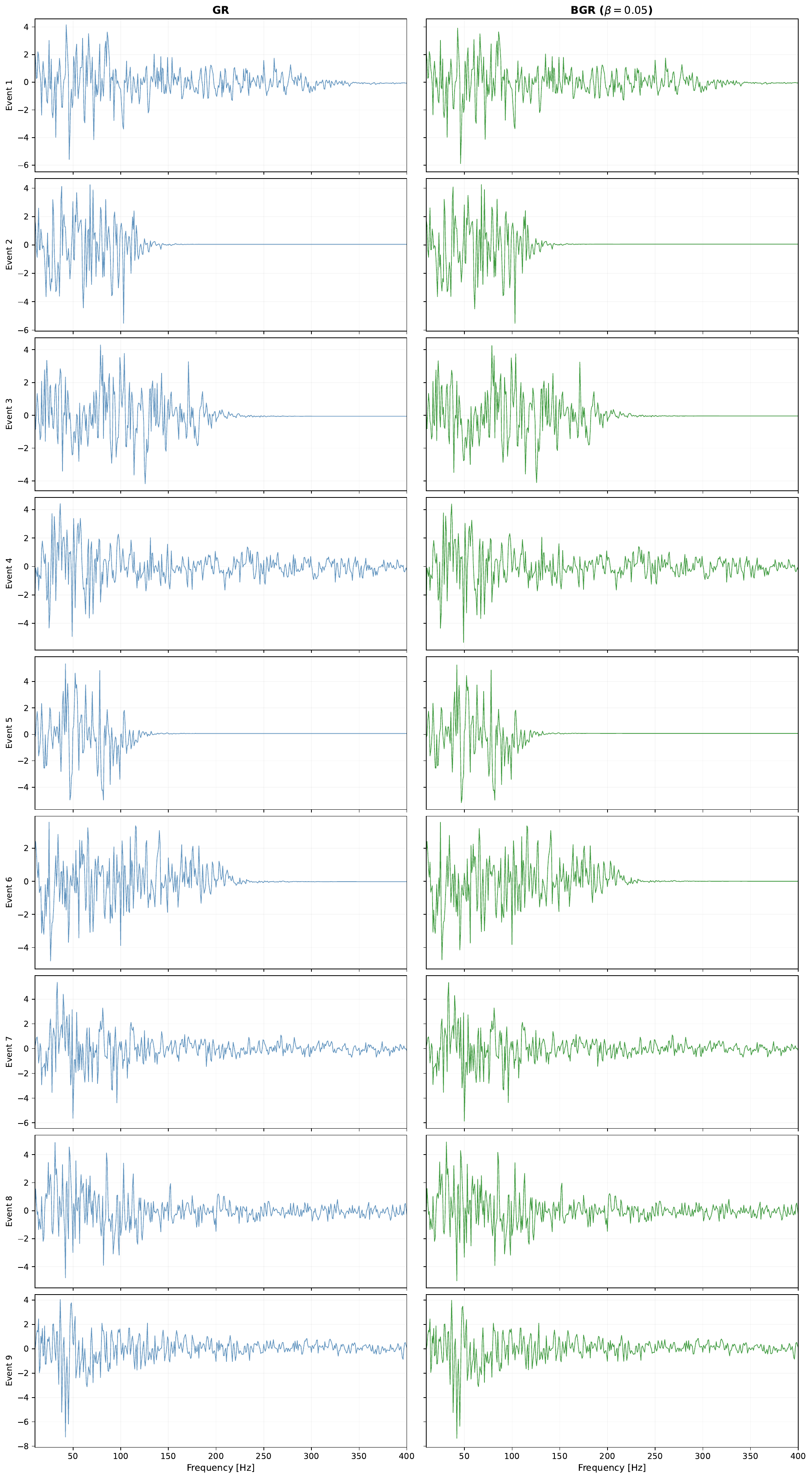}
    \caption{Normalized $\mathrm{Re}[R(f)]$ for nine events at $\beta = 0.05$. Left column: GR. Right column: BGR. Unlike the $\beta = 0.5$ example in Figure~\ref{fig:response_example}, the two cases are visually indistinguishable across all events, yet the CNN still achieves $58.9\%$ accuracy at this deformation scale.}
    \label{fig:response_small}
\end{figure}

To understand why the classification task becomes harder at small $\beta$, we visually examine individual response functions across different deformation scales. Figure~\ref{fig:response_example} in the previous section showed a clear difference between GR and BGR response functions at $\beta = 0.5$. Figure~\ref{fig:response_small} shows the same comparison at $\beta = 0.05$ for nine different events. At this smaller deformation, the GR and BGR response functions are visually indistinguishable across all events: both appear as noise-like fluctuations with no obvious difference in structure. Yet the CNN still achieves $58.9\%$ accuracy at $\beta = 0.05$, suggesting that the network captures subtle statistical patterns in the response function that are not visible on individual samples.

\begin{figure}[htbp]
    \centering
    \includegraphics[width=\textwidth]{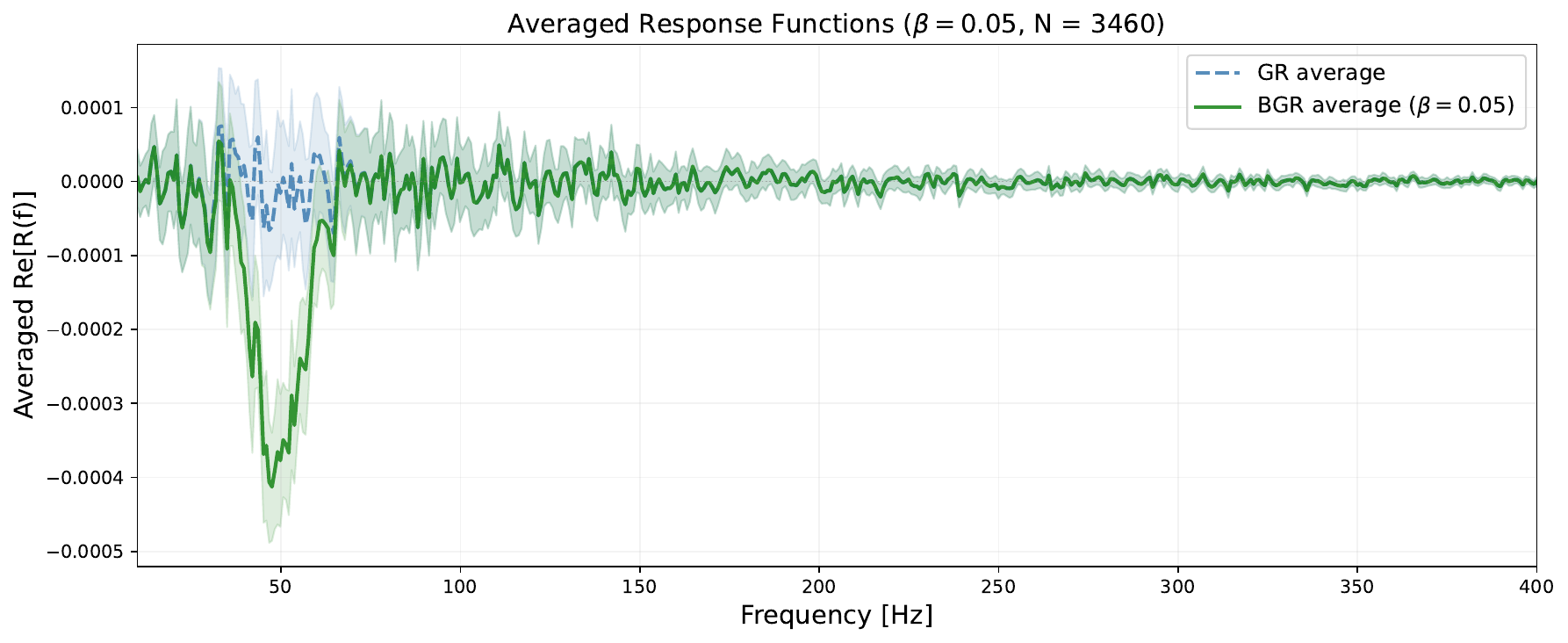}
    \caption{Averaged response functions at $\beta = 0.05$ over $N = 3{,}460$ samples. The GR average (blue, dashed) fluctuates around zero, while the BGR average (green, solid) shows a coherent dip near 50~Hz. The shaded bands indicate $\pm 1\sigma$ of the mean.}
    \label{fig:response_avg}
\end{figure}

Although the BGR signal is not visible in individual samples at small $\beta$, it can be revealed by averaging. When many BGR response functions from different events and noise realizations are averaged, the noise component (which has zero mean) cancels out, and only the coherent BGR signal survives. Figure~\ref{fig:response_avg} shows the averaged GR and BGR response functions at $\beta = 0.05$ over $N = 3{,}460$ samples. The GR average fluctuates around zero, while the BGR average develops a clear dip centered near 50~Hz, matching the location of the injected Gaussian deformation.

\begin{figure}[htbp]
    \centering
    \includegraphics[width=\textwidth]{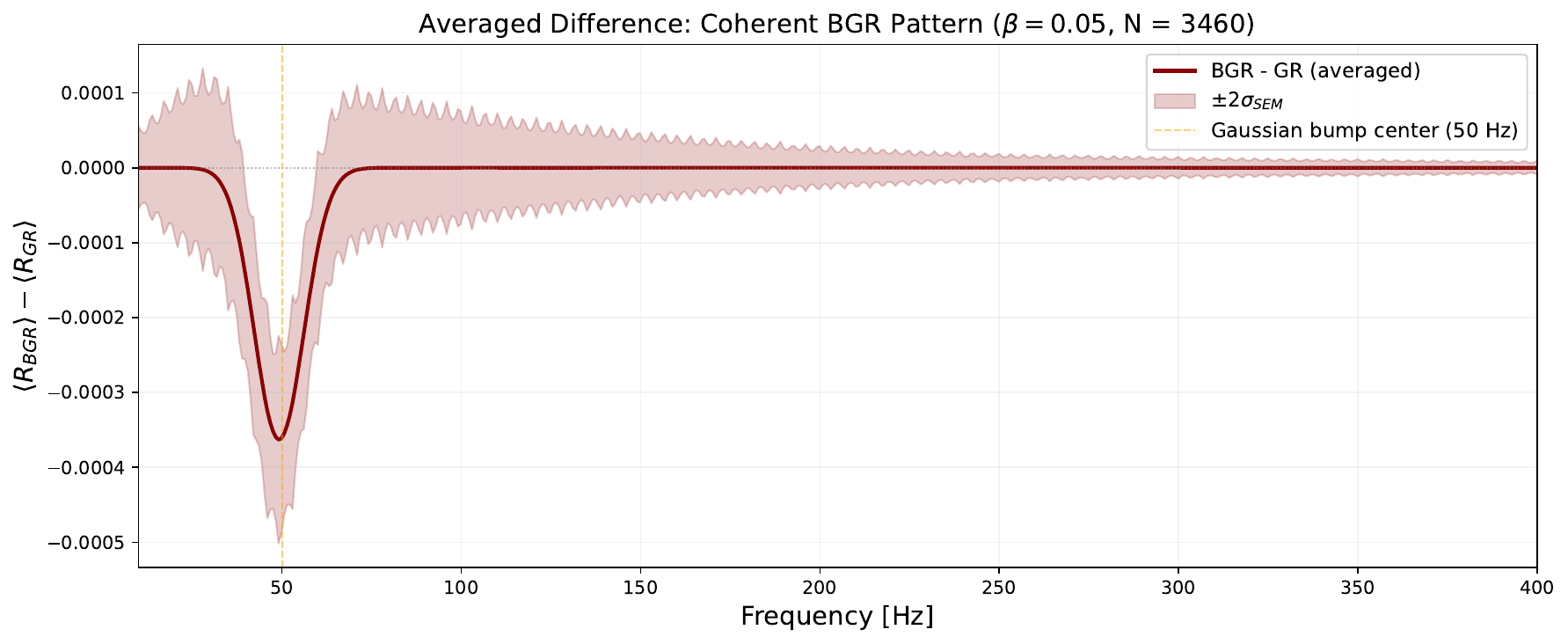}
    \caption{Difference between the averaged BGR and GR response functions, $\langle R_{\mathrm{BGR}} \rangle - \langle R_{\mathrm{GR}} \rangle$, at $\beta = 0.05$. The shaded band shows the $\pm 2\sigma_{\mathrm{SEM}}$ uncertainty. The dip centered at 50~Hz matches the shape of the injected phase deviation, confirming that the BGR signal is present even at this small deformation scale.}
    \label{fig:averaged_diff}
\end{figure}

Figure~\ref{fig:averaged_diff} shows the difference $\langle R_{\mathrm{BGR}} \rangle - \langle R_{\mathrm{GR}} \rangle$ at $\beta = 0.05$, isolating the coherent BGR pattern. The difference shows a Gaussian dip centered at 50~Hz with $\pm 2\sigma_{\mathrm{SEM}}$ error bands, matching the shape of the injected phase deviation $\delta\psi_{22}(f) = \exp[-(f-50)^2/100]$.

This confirms that the BGR signal is genuinely present in the response functions even at very small deformation scales, but is buried under noise in individual samples.

It is important to emphasize that the averaging method is a population level diagnostic, not a classification method. Computing $\langle R_{\mathrm{BGR}} \rangle - \langle R_{\mathrm{GR}} \rangle$ requires knowing in advance which samples are GR and which are BGR, which is precisely the information that a classifier is trying to infer. In a real detection scenario, one receives a single gravitational wave event and must decide whether it is consistent with GR or not, without access to an ensemble of labeled samples. The CNN operates in this single sample regime: it takes one response function and outputs a classification. The averaging result shows that the BGR signal exists at all $\beta$, but extracting it from a single noisy sample is a fundamentally harder problem than recovering it from a labeled population average. The gap between the two is not a failure of the CNN but a consequence of the Bayes optimal error (Section~\ref{sec:bayes_optimal}), which sets a fundamental lower bound on the misclassification rate when classifying individual samples from overlapping distributions.

\subsection{Bayes Optimal Error and the Classification Hierarchy}
\label{sec:bayes_optimal}

The CNN accuracy decreases as $\beta$ gets smaller, but it is natural to ask whether this is a limitation of our particular network or a fundamental property of the problem. The Bayes optimal error $\varepsilon_{\mathrm{Bayes}}$~\cite{devroye_probabilistic_1996,bishop_pattern_2006} provides a framework for thinking about this question. It is defined as the lowest possible classification error achievable by any classifier, given the overlap between the class conditional distributions. For a binary classification problem with equal class priors:
\begin{equation}
    \varepsilon_{\mathrm{Bayes}} = \mathbb{E}_{\mathbf{x}}\!\left[1 - \max_y P(y \mid \mathbf{x})\right],
\end{equation}
where $\mathbf{x}$ is the input feature vector and $y \in \{\mathrm{GR}, \mathrm{BGR}\}$ is the class label. When the GR and BGR distributions fully overlap ($\beta \to 0$), the posterior probabilities become $P(\mathrm{GR}|\mathbf{x}) = P(\mathrm{BGR}|\mathbf{x}) = 0.5$ for all $\mathbf{x}$, and $\varepsilon_{\mathrm{Bayes}} = 0.5$ (random chance). When the distributions are completely separated (large $\beta$), $\varepsilon_{\mathrm{Bayes}} = 0$ (perfect classification). Computing the Bayes error exactly requires knowing the full class conditional distributions over the entire input space, which is not feasible for our 500 dimensional response functions. However, we can build intuition by projecting the data down to a single dimension.

\begin{figure}[htbp]
    \centering
    \includegraphics[width=\textwidth]{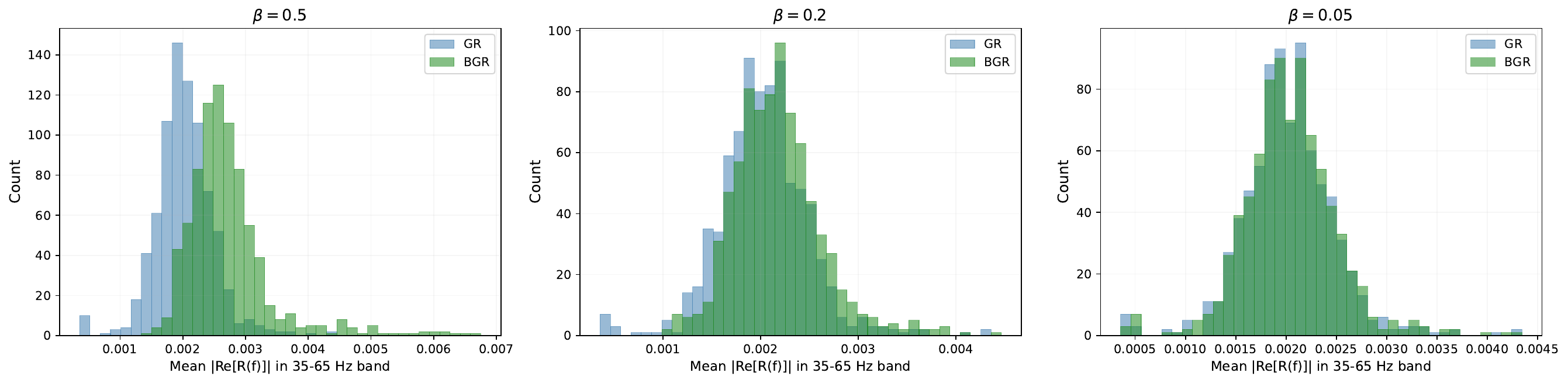}
    \caption{Distribution of the mean $|\mathrm{Re}[R(f)]|$ in the 35--65~Hz band for GR (blue) and BGR (green) samples at $\beta = 0.5$ (left), $\beta = 0.2$ (center), and $\beta = 0.05$ (right). As $\beta$ decreases, the distributions overlap more, increasing the Bayes error and making classification harder.}
    \label{fig:distribution_overlap}
\end{figure}

We reduce each response function to a single number: the mean of $|\mathrm{Re}[R(f)]|$ in the 35--65~Hz band, which is the frequency range around the injected Gaussian deformation. If a BGR response function carries extra structure near 50~Hz, this number will be systematically larger than for a GR response function, which contains only noise. Figure~\ref{fig:distribution_overlap} shows the histogram of this scalar feature for GR and BGR samples at three values of $\beta$.

At $\beta = 0.5$, the GR and BGR distributions are mostly separated, with the BGR distribution shifted to higher values due to the additional structured component in the response function. A simple threshold on this single number could already classify most samples correctly. At $\beta = 0.2$, the distributions overlap significantly, meaning many GR and BGR samples have similar amplitude values and no threshold works well. At $\beta = 0.05$ they are nearly identical, and classification based on this feature alone is essentially impossible. This increasing overlap directly corresponds to the decreasing classification accuracy observed in Table~\ref{tab:rf_results}.

The overlap region in these histograms represents samples that cannot be correctly classified based on this single feature alone. It is important to note that this one dimensional overlap is not the true Bayes error, which is defined over the full input space. The CNN operates on all 500 frequency bins simultaneously and may separate samples that appear indistinguishable in this one dimensional projection. Nevertheless, the histograms illustrate qualitatively why classification becomes harder at small $\beta$: the GR and BGR distributions converge, leaving less information for any classifier to exploit.

\subsection{CNN vs Single Feature Classifier}
\label{sec:cnn_vs_human}

The histograms in Section~\ref{sec:bayes_optimal} show that a single number, the mean amplitude near 50~Hz, already carries some information about the GR/BGR distinction. A natural question is whether the CNN does anything beyond detecting this one obvious feature. This particular feature is not an arbitrary choice: it is the most physically informed single number one could construct, since we know the deformation is centered at 50~Hz and the response function is designed to isolate the effect of phase deviations. Knowing that the deformation is centered at 50~Hz, one would naturally check this feature first when visually inspecting a response function for signs of a BGR signal.

\begin{figure}[htbp]
    \centering
    \includegraphics[width=0.9\textwidth]{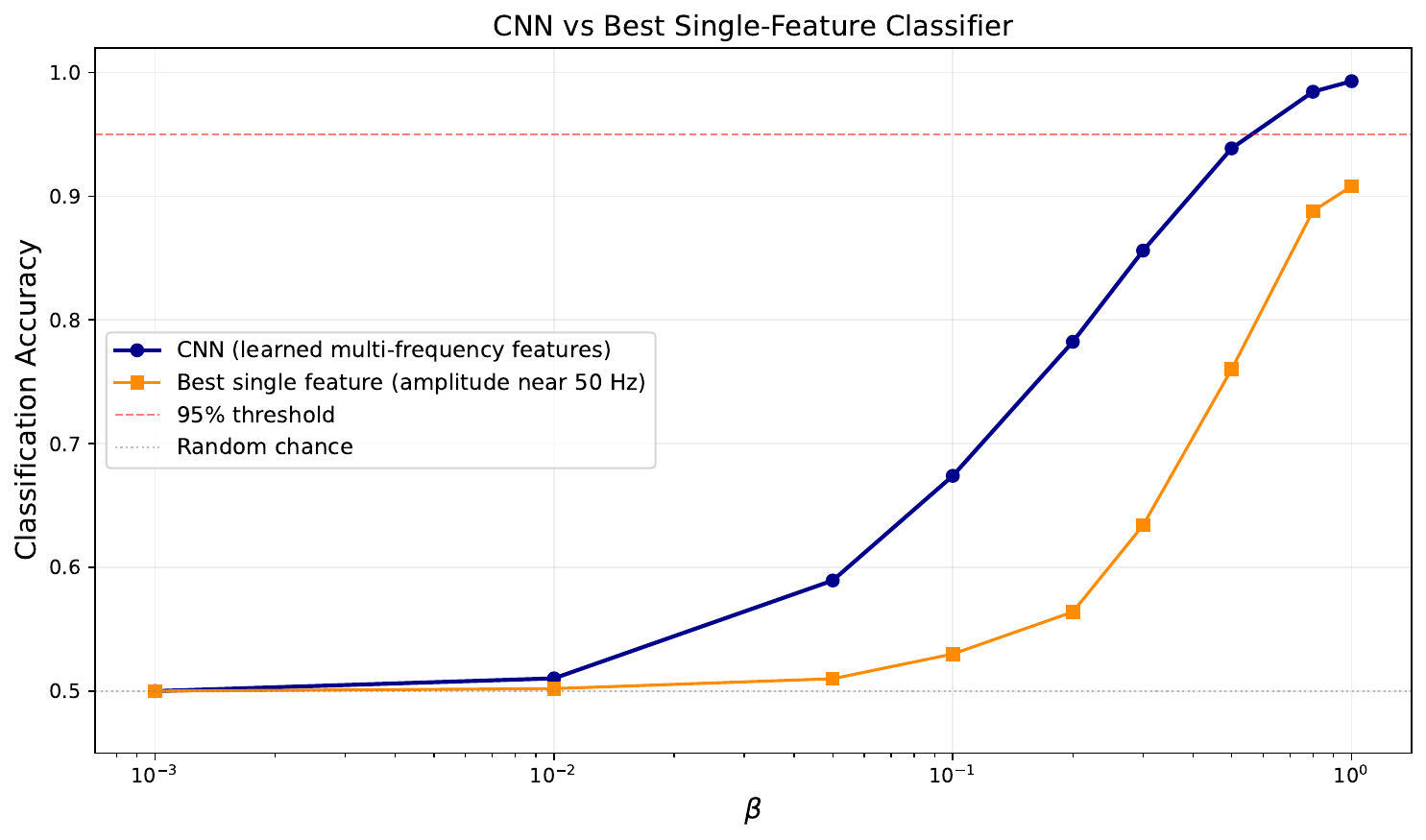}
    \caption{Classification accuracy of the CNN (blue circles) compared to the optimal single feature classifier based on the mean amplitude near 50~Hz (orange squares) as a function of $\beta$. The CNN consistently outperforms the single feature approach, indicating that it exploits correlations across multiple frequency bins.}
    \label{fig:cnn_vs_human}
\end{figure}

To test whether the CNN goes beyond this, we build the best possible classifier that uses only this single number. We scan over all possible threshold values on the training set and select the one that maximizes accuracy. A sample is classified as BGR if its mean amplitude exceeds this threshold and as GR otherwise. Since the threshold is chosen optimally on the training set, this represents the best possible classification using only this single feature. If the CNN only matched this classifier, it would mean the network is simply learning to detect the amplitude bump near 50~Hz and nothing more. Figure~\ref{fig:cnn_vs_human} compares the CNN accuracy to this single feature classifier across all $\beta$ values.

At $\beta = 1.0$, the CNN achieves $99.3\%$ accuracy while the single feature classifier reaches $90.8\%$. At $\beta = 0.5$, the CNN achieves $93.9\%$ compared to $76.0\%$ for the single feature. At $\beta = 0.3$, the gap widens further: $85.6\%$ (CNN) vs $63.4\%$ (single feature). The CNN consistently outperforms the single feature classifier at every $\beta$ value.

The CNN consistently outperforms the single feature classifier, even though the latter uses the most informative single feature with an optimal threshold. This means the CNN extracts additional information from the shape and correlations across frequencies that cannot be captured by any single number, no matter how well chosen. This is something that cannot be done by visual inspection of individual response functions.


\section{Parameterized Post-Einsteinian Classification}
\label{sec:ppe}
\label{sec:ppe_framework}

The Gaussian toy model studied in the previous sections provides a controlled way to probe the sensitivity of our classification framework, but it is not derived from any specific modified gravity theory. In this section, we move toward physically motivated deformations using the parameterized post-Einsteinian (ppE) framework~\cite{yunes_fundamentaltheoreticalbiasgravitationalwaveastrophysicsparameterizedposteinsteinianframework_2009,tahura_parameterizedposteinsteiniangravitationalwaveformsvariousmodifiedtheoriesgravity_2018}. The ppE framework provides a systematic way to parameterize deviations from GR as power law phase corrections, where each modified gravity theory maps onto a specific exponent. We first review the framework and the ppE signatures of several representative theories, and then apply the classification pipeline to massive gravity as a concrete demonstration.

\subsection{The ppE Framework and Modified Gravity Theories}
\label{sec:ppe_theories}

In the stationary phase approximation, the GR waveform in the frequency domain takes the form
\begin{equation}
    \tilde{h}_{\mathrm{GR}}(f) = \mathcal{A}_{\mathrm{GR}}(f)\, e^{i\Psi_{\mathrm{GR}}(f)}\,,
\end{equation}
where $\mathcal{A}_{\mathrm{GR}}(f)$ is the amplitude and $\Psi_{\mathrm{GR}}(f)$ is the phase. The ppE framework introduces parametric deviations in both amplitude and phase~\cite{yunes_fundamentaltheoreticalbiasgravitationalwaveastrophysicsparameterizedposteinsteinianframework_2009}:
\begin{equation}
    \tilde{h}_{\mathrm{ppE}}(f) = \tilde{h}_{\mathrm{GR}}(f)\left(1 + \alpha\, u^a\right) e^{i\beta\, u^b}\,,
    \label{eq:ppe_waveform}
\end{equation}
where $u = \pi M f$ is the reduced frequency, and the parameters $(\alpha, a)$ and $(\beta, b)$ characterize the deviations in amplitude and phase respectively. GR is recovered when $\alpha = \beta = 0$.

Since phase modifications accumulate over many orbital cycles and are generally more detectable than amplitude changes~\cite{mehta_testsgeneralrelativitygravitationalwaveobservationsusingflexibletheoryindependentmethod_2023}, we focus exclusively on phase-only deviations by setting $\alpha = 0$:
\begin{equation}
    \tilde{h}_{\mathrm{ppE}}(f) = \tilde{h}_{\mathrm{GR}}(f)\, e^{i\beta\, u^b}\,.
    \label{eq:ppe_phase_only}
\end{equation}
The phase correction is then
\begin{equation}
    \delta\psi(f) = \beta\, (\pi M f)^b\,,
    \label{eq:ppe_delta_psi}
\end{equation}
which has the same structure as our general framework (Section~\ref{sec:param_deviations}), but with a specific power law form determined by the theory. Different values of $b$ correspond to different post-Newtonian orders, and different modified theories of gravity predict specific values of $(\beta, b)$~\cite{tahura_parameterizedposteinsteiniangravitationalwaveformsvariousmodifiedtheoriesgravity_2018,Maggio:2022hre,Pompili:2025cdc,Piarulli:2025rvr}. We briefly describe three representative examples.

\paragraph{Brans-Dicke theory.} Brans-Dicke theory~\cite{brans_machsprinciple_1961} is a scalar tensor theory that promotes Newton's constant to a dynamical scalar field coupled to the Ricci scalar with a coupling constant $\omega_{\mathrm{BD}}$. The theory predicts dipole radiation from binary systems, which introduces a $-1$PN correction to the phase:
\begin{equation}
    \beta_{\mathrm{BD}} = -\frac{5}{3584\, \omega_{\mathrm{BD}}}\, (s_1 - s_2)^2\, \eta^{2/5}\,, \qquad b = -\frac{7}{3}\,,
\end{equation}
where $s_1$ and $s_2$ are the scalar charges of the two compact objects. However, in standard Brans-Dicke theory, black holes do not carry scalar charge due to no-hair theorems~\cite{hawking_blackholesbransdicke_1972,sotiriou_blackholesscalartensorgravity_2012}, so $\beta_{\mathrm{BD}} = 0$ for binary black hole systems.

\paragraph{Massive gravity.} In massive graviton theories~\cite{rham_massivegravity_2014,hinterbichler_theoreticalaspectsmassivegravity_2012,will_gravitonmass_1998}, the graviton has a small but nonzero mass $m_g$, which modifies the dispersion relation of gravitational waves. Higher frequency waves travel faster than lower frequency ones, producing a cumulative phase shift that grows with the source distance $D$:
\begin{equation}
    \beta_{\mathrm{MG}} = \frac{\pi^2 D \mathcal{M}}{\lambda_g^2 (1 + z)}\,, \qquad b = -1\,,
\end{equation}
where $\lambda_g = h/(m_g c)$ is the graviton Compton wavelength, $\mathcal{M} = \eta^{3/5} M$ is the chirp mass, and $z$ is the source redshift. The phase correction is
\begin{equation}
    \delta\psi_{\mathrm{MG}}(f) = -\beta_{\mathrm{MG}}\, (\pi \mathcal{M} f)^{-1}\,,
    \label{eq:mg_phase}
\end{equation}
which enters at $+1$PN order and vanishes in the limit $m_g \to 0$.

\paragraph{Dynamical Chern-Simons gravity.} Dynamical Chern-Simons (dCS) gravity~\cite{jackiw_chernsimonsmodification_2003,alexander_chernsimonsreview_2009,canizares_testingchernsimonsmodifiedgravitygravitationalwavedetectionsextrememassratiobinaries_2012} introduces a parity-violating coupling between a dynamical scalar field and the gravitational Pontryagin density. Spinning black holes acquire scalar charge in this theory, leading to a phase correction at fractional PN order:
\begin{equation}
    \beta_{\mathrm{dCS}} \propto \zeta_{\mathrm{CS}} \left(\frac{m_1^2}{M^2}\chi_1^2 + \frac{m_2^2}{M^2}\chi_2^2\right), \qquad b = -\frac{1}{3}\,,
\end{equation}
where $\zeta_{\mathrm{CS}} \propto \xi/M^4$ is a dimensionless coupling parameter.

Each of these theories produces a phase correction of the form $\delta\psi(f) = \beta\, (\pi M f)^b$ with a distinct value of $b$, providing different frequency dependent signatures. Among these, massive gravity is the most natural candidate for our binary black hole dataset: unlike Brans-Dicke, it affects BBH systems, and unlike dCS, the phase shift depends on the source distance, making use of the realistic luminosity distances in our catalog. We therefore demonstrate the classification framework on massive gravity in the following subsection. The extension to other theories, including dCS, follows the same procedure and is left for future work.

\subsection{Binary Classification of Massive Gravity Waveforms}
\label{sec:ppe_binary}

We now apply the response function classification framework to massive gravity. For each injection in our dataset, we compute the ppE phase correction from Eq.~\eqref{eq:mg_phase} using its chirp mass $\mathcal{M}$, luminosity distance $D$, and redshift $z$. This means the strength of the phase shift varies from injection to injection, reflecting the physical dependence of the massive gravity effect on the source properties and distance.

We construct BGR waveforms by inserting the massive gravity phase correction into the exact BGR formula (Eq.~\eqref{eq:bgr_exact}), using $\delta\psi_{22}(f) = -\beta_{\mathrm{MG}}\, (\pi \mathcal{M} f)^{-1}$, and then compute response functions as described in Section~\ref{sec:mismatch_response}. The same CNN architecture, training procedure, and event level split are used.

\begin{figure}[htbp]
    \centering
    \includegraphics[width=\textwidth]{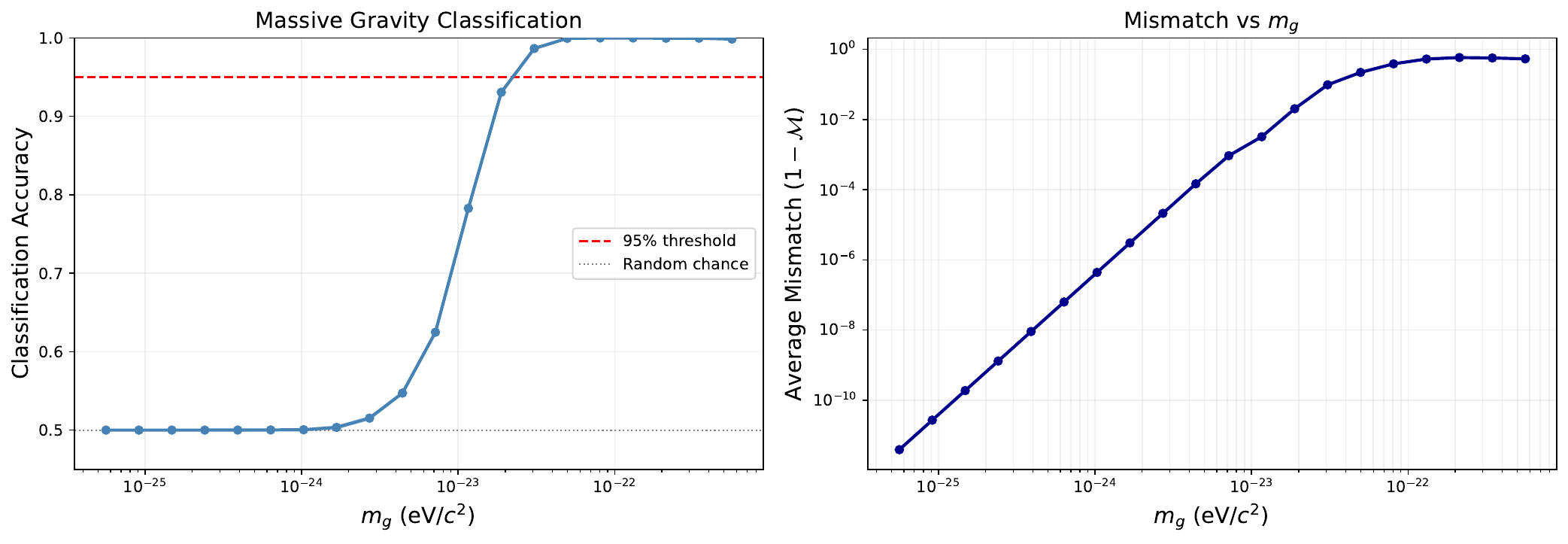}
    \caption{Left: classification accuracy as a function of the graviton mass $m_g$ using response function classification. The 95\% threshold (dashed red) is crossed at $m_g \approx 2 \times 10^{-23}$~eV/$c^2$. Right: average waveform mismatch as a function of $m_g$. The mismatch increases with $m_g$ and saturates above $\sim 10^{-22}$~eV/$c^2$.}
    \label{fig:mg_summary}
\end{figure}

We sweep over graviton masses from $m_g \sim 6 \times 10^{-26}$~eV/$c^2$ to $m_g \sim 6 \times 10^{-22}$~eV/$c^2$ and evaluate the classifier's accuracy at each value. Figure~\ref{fig:mg_summary} shows the classification accuracy and the corresponding waveform mismatch as a function of $m_g$.

The classifier crosses the 95\% accuracy threshold at a graviton mass of approximately
\begin{equation}
    m_g \approx 2 \times 10^{-23}\;\mathrm{eV}/c^2\,.
\end{equation}
Below this value, the accumulated phase shift becomes too small for the classifier to reliably distinguish massive gravity waveforms from GR predictions. Above this threshold, the accuracy rises steeply, reaching $98.7\%$ at $m_g \approx 3 \times 10^{-23}$~eV/$c^2$ and $99.9\%$ at $m_g \approx 5 \times 10^{-23}$~eV/$c^2$.

This demonstrates that the response function classification pipeline, developed and validated on the Gaussian toy model in the previous sections, applies directly to physically motivated theories. The classification threshold of $m_g \approx 2 \times 10^{-23}$~eV/$c^2$ is of the same order as current observational bounds~\cite{abbott_tests_2021,LIGOScientific:2026fcf,rham_gravitonmassbounds_2017}, showing that the framework probes the physically relevant regime for massive gravity.

\subsection{Toward Multi-class Theory Identification}
\label{sec:ppe_multiclass}

\begin{figure}[htbp]
    \centering
    \includegraphics[width=\textwidth]{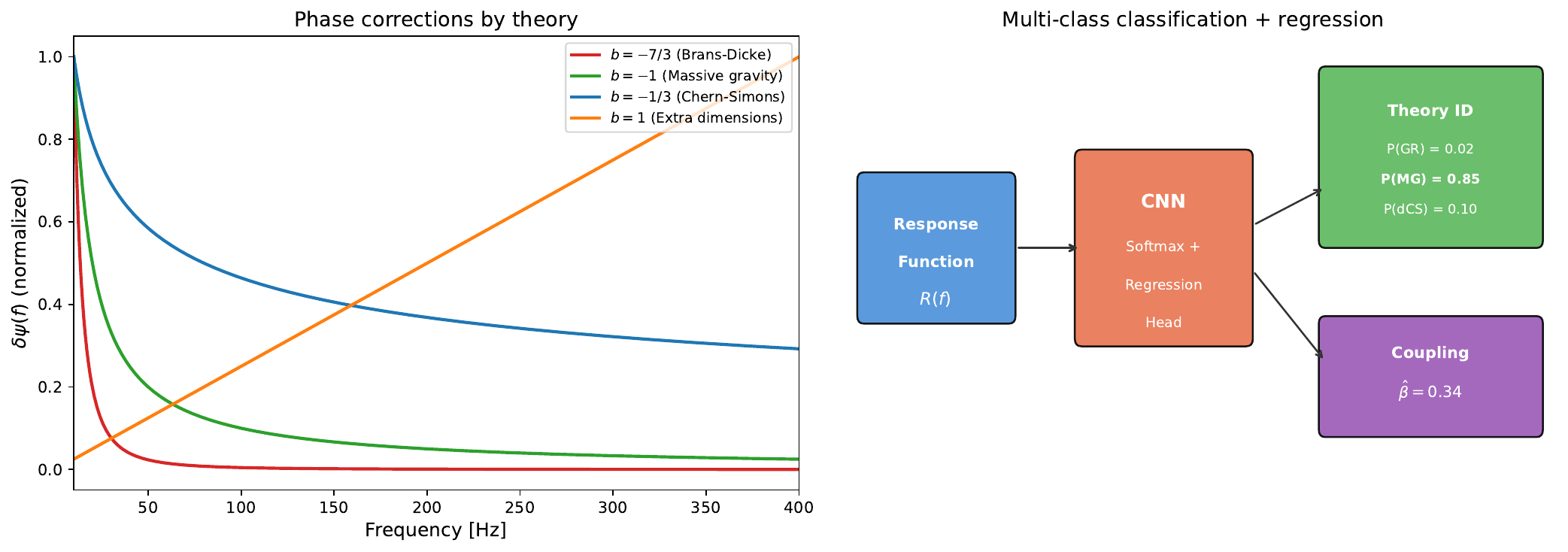}
    \caption{Left: normalized phase corrections $\delta\psi(f)$ for four representative ppE theories with different exponents $b$. Each theory produces a distinct frequency dependent signature. Right: schematic of the multi-class classification and regression pipeline. The response function $R(f)$ is fed into a CNN with a softmax head for theory identification and a regression head for coupling strength estimation.}
    \label{fig:multiclass_schematic}
\end{figure}

The binary classification framework developed in this work answers a single question: is this event consistent with GR or not? A more ambitious goal is to identify \emph{which} modified gravity theory produced a deviation, and to estimate its coupling strength. The ppE framework provides a natural setting for this, since each theory predicts a phase correction $\delta\psi(f) = \beta\, (\pi M f)^b$ with a distinct exponent $b$. As shown in the left panel of Figure~\ref{fig:multiclass_schematic}, the different values of $b$ produce phase corrections with very different frequency dependence: theories with negative $b$ (such as Brans-Dicke and massive gravity) concentrate the deviation at low frequencies, while theories with positive $b$ (such as extra-dimensional models) grow at high frequencies. These distinct spectral shapes are precisely the kind of features that a CNN can learn to distinguish.

The right panel of Figure~\ref{fig:multiclass_schematic} illustrates how this could work in practice. The same response function $R(f)$ used for binary classification serves as input to a CNN with two output heads: a softmax classification head that assigns probabilities to each candidate theory (including GR), and a regression head that estimates the coupling strength $\beta$. In a single forward pass, the network would output both the most likely theory and the estimated deformation strength. This is a task where the CNN approach may offer a genuine advantage over Bayesian parameter estimation, which requires running separate analyses for each candidate theory. A detailed study of multi-class theory identification and coupling regression using the response function framework is currently a work in progress.

\section{Conclusion}
\label{sec:conclusion}

In this work, we developed a machine learning framework for testing general relativity using simulated gravitational wave signals from binary black hole mergers. Using the source parameters of 173 BBH events from the GWTC catalog as a realistic astrophysical population, we trained convolutional neural networks to distinguish GR waveforms from beyond GR waveforms constructed through a controlled Gaussian phase deformation of the dominant $(2,2)$ mode.

We compared two input representations for the classifier. The first uses the whitened waveforms directly, providing the network with the full frequency domain signal content. The second uses the response function $R(f)$, obtained by applying the response function formalism to the waveform mismatch, which effectively divides out the noise contribution and isolates the deviation from GR. Response functions improve the classification sensitivity by more than one order of magnitude compared to whitened waveforms, demonstrating that the choice of observable representation is as important as the classifier architecture itself.

We investigated the fundamental limits of the classification task through several complementary approaches. Bayes optimal error analysis provides a lower bound on the achievable classification error. Visual diagnostics through averaging methods reveal that coherent patterns emerge in the response functions when many events are combined, even at deformation strengths where individual event classification fails. A comparison between CNN performance and an optimal single feature classifier confirms that the network extracts information across the full frequency range rather than relying on a single localized spectral feature.

We connected the framework to physically motivated theories through the parameterized post Einsteinian (ppE) formalism and applied it to massive gravity as a concrete demonstration. The classifier reaches $95\%$ accuracy at a graviton mass of $m_g \approx 2 \times 10^{-23}\;\mathrm{eV}/c^2$ with aLIGO design sensitivity, which is of the same order as current observational bounds~\cite{abbott_tests_2021,LIGOScientific:2026fcf}.

The standard approach to testing GR with gravitational wave data is Bayesian parameter estimation~\cite{abbott_tests_2021,LIGOScientific:2026qni,LIGOScientific:2026fcf}, in which ppE deviation parameters are included alongside the standard binary parameters and their posteriors are checked for consistency with GR. Our classification approach is not a replacement for this framework but explores a complementary question: whether a CNN can learn to detect deviations directly from the frequency domain representation without explicit template matching.

Where the CNN approach may offer a structural advantage is in the multiclass setting. In the Bayesian framework, testing multiple modified gravity theories requires running separate parameter estimation analyses for each candidate theory (each value of the ppE exponent $b$), followed by pairwise Bayes factor computations. A multiclass CNN, by contrast, takes the response function as input and outputs a probability distribution over all candidate theories in a single forward pass. This makes the question ``which theory produced the deviation?'' naturally tractable without scaling the computational effort with the number of theories. Multiclass theory identification and coupling strength regression using the response function framework is currently a work in progress.


\acknowledgments
We acknowledge the funding from the European Research Council (ERC) under the European Unions Horizon 2020 research and innovation programme grant agreement No 801781. L.H. further acknowledges support from the Deutsche Forschungsgemeinschaft (DFG, German Research Foundation)
under Germany’s Excellence Strategy EXC 2181/1 - 390900948 (the Heidelberg STRUCTURES
Excellence Cluster).


\appendix

\section{Neural Network Architecture}
\label{app:nn_architecture}

The convolutional neural network used throughout this work takes as input a two channel frequency domain representation of shape $(500, 2)$, where the two channels correspond to the real and imaginary parts of the whitened observable (either the whitened waveform or the response function). The frequency range is $10$--$400$~Hz, sampled at 500 equally spaced points.

The architecture consists of three convolutional blocks followed by a dense classification head. Each convolutional block contains a one dimensional convolution (Conv1D), batch normalization, a ReLU activation, spatial dropout (SpatialDropout1D), and a pooling layer. The number of filters increases across blocks (32, 64, 128) while the kernel size decreases (11, 7, 5), allowing the network to capture both broad and fine scale features in the frequency domain. The first two blocks use max pooling with pool size 2, while the third block uses global average pooling to reduce the feature maps to a single vector per filter. L2 weight regularization with coefficient $5 \times 10^{-4}$ is applied to all convolutional and dense layers.

\begin{figure}[h]
\centering
\includegraphics[width=\textwidth]{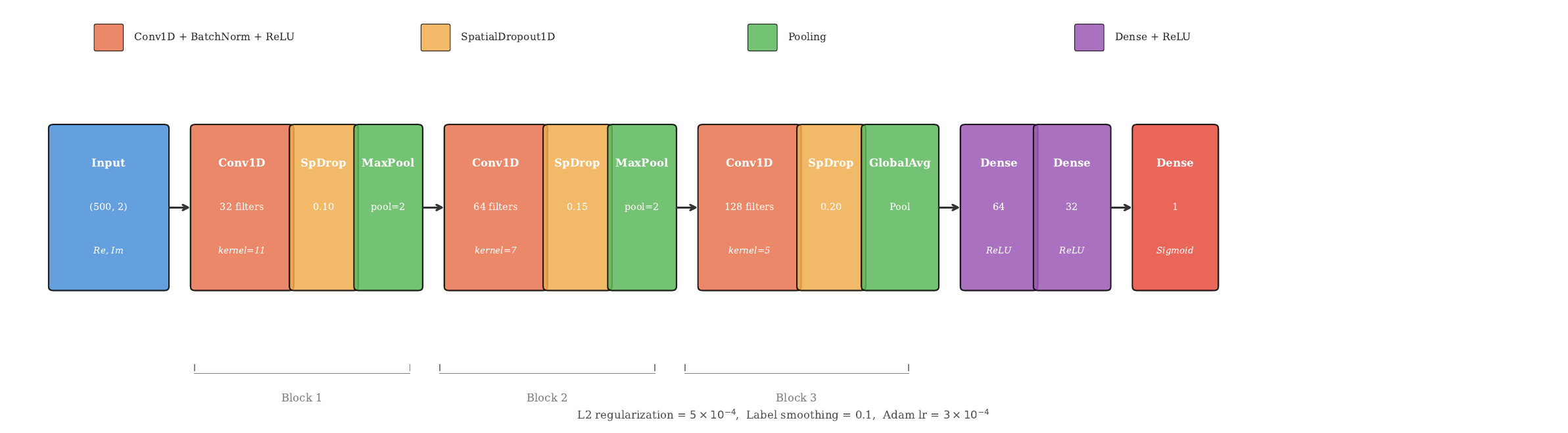}
\caption{Schematic diagram of the CNN architecture. The network consists of three convolutional blocks with increasing filter counts (32, 64, 128) and decreasing kernel sizes (11, 7, 5), followed by a dense classification head. Each convolutional block includes batch normalization, ReLU activation, spatial dropout, and pooling.}
\label{fig:cnn_architecture}
\end{figure}

The dense head consists of two fully connected layers with 64 and 32 units respectively, each followed by ReLU activation. The output layer has a single unit with a sigmoid activation for binary classification. The full architecture is summarized in Table~\ref{tab:architecture} and illustrated schematically in Figure~\ref{fig:cnn_architecture}.

\begin{table}[h]
\centering
\begin{tabular}{llcc}
\toprule
\textbf{Layer} & \textbf{Type} & \textbf{Output Shape} & \textbf{Parameters} \\
\midrule
Input & --- & $(500, 2)$ & 0 \\
\midrule
\multicolumn{4}{l}{\textit{Convolutional Block 1}} \\
Conv1D & 32 filters, kernel 11 & $(500, 32)$ & 736 \\
BatchNorm & --- & $(500, 32)$ & 128 \\
ReLU & --- & $(500, 32)$ & 0 \\
SpatialDropout1D & rate 0.10 & $(500, 32)$ & 0 \\
MaxPool1D & pool size 2 & $(250, 32)$ & 0 \\
\midrule
\multicolumn{4}{l}{\textit{Convolutional Block 2}} \\
Conv1D & 64 filters, kernel 7 & $(250, 64)$ & 14{,}400 \\
BatchNorm & --- & $(250, 64)$ & 256 \\
ReLU & --- & $(250, 64)$ & 0 \\
SpatialDropout1D & rate 0.15 & $(250, 64)$ & 0 \\
MaxPool1D & pool size 2 & $(125, 64)$ & 0 \\
\midrule
\multicolumn{4}{l}{\textit{Convolutional Block 3}} \\
Conv1D & 128 filters, kernel 5 & $(125, 128)$ & 41{,}088 \\
BatchNorm & --- & $(125, 128)$ & 512 \\
ReLU & --- & $(125, 128)$ & 0 \\
SpatialDropout1D & rate 0.20 & $(125, 128)$ & 0 \\
GlobalAvgPool1D & --- & $(128,)$ & 0 \\
\midrule
\multicolumn{4}{l}{\textit{Dense Head}} \\
Dense & 64 units, ReLU & $(64,)$ & 8{,}256 \\
Dense & 32 units, ReLU & $(32,)$ & 2{,}080 \\
Dense & 1 unit, Sigmoid & $(1,)$ & 33 \\
\midrule
\multicolumn{2}{l}{\textbf{Total parameters}} & & \textbf{67{,}489} \\
\bottomrule
\end{tabular}
\caption{Summary of the CNN architecture. All convolutional and dense layers use L2 regularization with coefficient $5 \times 10^{-4}$. The same architecture is used for both waveform and response function classification.}
\label{tab:architecture}
\end{table}

\begin{figure}[h]
\centering
\includegraphics[width=\textwidth]{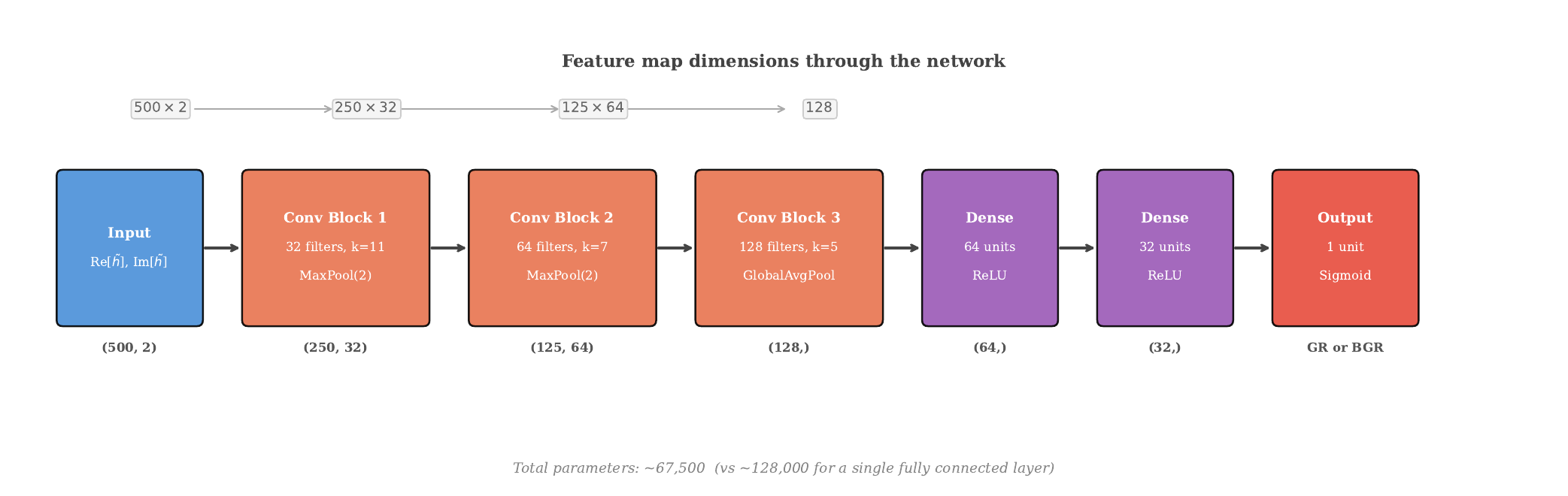}
\caption{Data flow and feature map dimensions through the CNN. The input $(500, 2)$ is progressively transformed through three convolutional blocks with pooling, reducing the frequency dimension while increasing the number of feature maps. The dense head maps the resulting 128 dimensional vector to a single output for binary classification.}
\label{fig:cnn_flow}
\end{figure}

Figure~\ref{fig:cnn_flow} shows the data flow through the network together with the feature map dimensions at each stage. The frequency dimension is progressively reduced from 500 to 250 to 125 through max pooling, while the number of learned feature maps increases from 2 to 32 to 64 to 128. Global average pooling then collapses the spatial dimension entirely, producing a 128 dimensional feature vector that is passed to the dense classification head.

The network is trained using the Adam optimizer~\cite{kingma_adam_2015} with a learning rate of $3 \times 10^{-4}$ and binary cross entropy loss with label smoothing of $0.1$~\cite{szegedy_rethinking_2016}. Training runs for up to 80 epochs with early stopping based on the validation loss (patience of 10 epochs). For each classification task and deformation strength, we train 4 independent models with different random seeds and report the mean and standard deviation of the test accuracy.

\section{Evaluation Metrics}
\label{app:eval_metrics}

We evaluate the classifier using standard binary classification metrics. For a binary classifier with positive class BGR and negative class GR, we define:
\begin{itemize}[leftmargin=*]
    \item \textbf{True Positives (TP):} BGR samples correctly classified as BGR.
    \item \textbf{True Negatives (TN):} GR samples correctly classified as GR.
    \item \textbf{False Positives (FP):} GR samples incorrectly classified as BGR.
    \item \textbf{False Negatives (FN):} BGR samples incorrectly classified as GR.
\end{itemize}

The \textbf{accuracy} is the fraction of all samples that are correctly classified:
\begin{equation}
\mathrm{Accuracy} = \frac{\mathrm{TP} + \mathrm{TN}}{\mathrm{TP} + \mathrm{TN} + \mathrm{FP} + \mathrm{FN}}\,.
\end{equation}

The \textbf{precision} measures what fraction of samples classified as BGR are actually BGR:
\begin{equation}
\mathrm{Precision} = \frac{\mathrm{TP}}{\mathrm{TP} + \mathrm{FP}}\,.
\end{equation}

The \textbf{recall} (also called sensitivity or true positive rate) measures what fraction of actual BGR samples are correctly identified:
\begin{equation}
\mathrm{Recall} = \frac{\mathrm{TP}}{\mathrm{TP} + \mathrm{FN}}\,.
\end{equation}

The \textbf{F1 score} is the harmonic mean of precision and recall, providing a single metric that balances both:
\begin{equation}
\mathrm{F1} = 2 \cdot \frac{\mathrm{Precision} \cdot \mathrm{Recall}}{\mathrm{Precision} + \mathrm{Recall}}\,.
\end{equation}

In this work, we use balanced datasets with equal numbers of GR and BGR samples, so accuracy, precision, recall, and F1 score are closely related. We report accuracy as the primary metric throughout the paper, and show precision, recall, and F1 in the confusion matrices (Section~\ref{sec:waveform_results}) where the per class performance is relevant.


\bibliographystyle{JHEP}
\bibliography{biblio.bib}

@article{abbott_observation_2016,
    author = "{Abbott, B. P. and others}",
    collaboration = "LIGO Scientific, Virgo",
    title = "{Observation of Gravitational Waves from a Binary Black Hole Merger}",
    journal = "Phys. Rev. Lett.",
    volume = "116",
    pages = "061102",
    year = "2016",
    doi = "10.1103/PhysRevLett.116.061102",
    eprint = "1602.03837",
    archivePrefix = "arXiv",
    primaryClass = "gr-qc"
}

@article{abbott_gwtc1_2019,
    author = "{Abbott, B. P. and others}",
    collaboration = "LIGO Scientific, Virgo",
    title = "{GWTC-1: A Gravitational-Wave Transient Catalog of Compact Binary Mergers Observed by LIGO and Virgo during the First and Second Observing Runs}",
    journal = "Phys. Rev. X",
    volume = "9",
    pages = "031040",
    year = "2019",
    doi = "10.1103/PhysRevX.9.031040",
    eprint = "1811.12907",
    archivePrefix = "arXiv",
    primaryClass = "astro-ph.HE"
}

@article{abbott_gwtc2_2021,
    author = "{Abbott, R. and others}",
    collaboration = "LIGO Scientific, Virgo",
    title = "{GWTC-2: Compact Binary Coalescences Observed by LIGO and Virgo during the First Half of the Third Observing Run}",
    journal = "Phys. Rev. X",
    volume = "11",
    pages = "021053",
    year = "2021",
    doi = "10.1103/PhysRevX.11.021053",
    eprint = "2010.14527",
    archivePrefix = "arXiv",
    primaryClass = "gr-qc"
}

@article{abbott_gwtc3_2023,
    author = "{Abbott, R. and others}",
    collaboration = "LIGO Scientific, Virgo, KAGRA",
    title = "{GWTC-3: Compact Binary Coalescences Observed by LIGO and Virgo during the Second Part of the Third Observing Run}",
    journal = "Phys. Rev. X",
    volume = "13",
    pages = "041039",
    year = "2023",
    doi = "10.1103/PhysRevX.13.041039",
    eprint = "2111.03606",
    archivePrefix = "arXiv",
    primaryClass = "gr-qc"
}

@article{abbott_tests_2021,
    author = "{Abbott, R. and others}",
    collaboration = "LIGO Scientific, Virgo, KAGRA",
    title = "{Tests of General Relativity with GWTC-3}",
    journal = "Phys. Rev. D",
    volume = "112",
    pages = "084080",
    year = "2025",
    doi = "10.1103/PhysRevD.112.084080",
    eprint = "2112.06861",
    archivePrefix = "arXiv",
    primaryClass = "gr-qc"
}

@article{yunes_fundamentaltheoreticalbiasgravitationalwaveastrophysicsparameterizedposteinsteinianframework_2009,
    author = "Yunes, Nicol\'as and Pretorius, Frans",
    title = "{Fundamental Theoretical Bias in Gravitational Wave Astrophysics and the Parameterized Post-Einsteinian Framework}",
    journal = "Phys. Rev. D",
    volume = "80",
    pages = "122003",
    year = "2009",
    doi = "10.1103/PhysRevD.80.122003",
    eprint = "0909.3328",
    archivePrefix = "arXiv",
    primaryClass = "gr-qc"
}

@article{tahura_parameterizedposteinsteiniangravitationalwaveformsvariousmodifiedtheoriesgravity_2018,
    author = "Tahura, Shammi and Yagi, Kent",
    title = "{Parameterized Post-Einsteinian Gravitational Waveforms in Various Modified Theories of Gravity}",
    journal = "Phys. Rev. D",
    volume = "98",
    pages = "084042",
    year = "2018",
    doi = "10.1103/PhysRevD.98.084042",
    eprint = "1809.00259",
    archivePrefix = "arXiv",
    primaryClass = "gr-qc"
}

@article{xie_neuralposteinsteinianframeworkefficienttheoryagnostictestsgeneralrelativitygravitationalwaves_2024,
    author = "Xie, Yiqi and others",
    title = "{Neural Post-Einsteinian Framework for Efficient Theory-Agnostic Tests of General Relativity with Gravitational Waves}",
    journal = "Phys. Rev. D",
    volume = "110",
    pages = "024036",
    year = "2024",
    doi = "10.1103/PhysRevD.110.024036",
    eprint = "2403.18936",
    archivePrefix = "arXiv",
    primaryClass = "gr-qc"
}

@article{cuoco_enhancing_2021,
    author = "Cuoco, Elena and others",
    title = "{Enhancing Gravitational-Wave Science with Machine Learning}",
    journal = "Mach. Learn.: Sci. Technol.",
    volume = "2",
    pages = "011002",
    year = "2021",
    doi = "10.1088/2632-2153/abb93a",
    eprint = "2005.03745",
    archivePrefix = "arXiv",
    primaryClass = "astro-ph.IM"
}

@article{huerta_accelerated_2021,
    author = "Huerta, E. A. and others",
    title = "{Accelerated, Scalable and Reproducible AI-driven Gravitational Wave Detection}",
    journal = "Nature Astron.",
    volume = "5",
    pages = "1062--1068",
    year = "2021",
    doi = "10.1038/s41550-021-01405-0",
    eprint = "2012.08545",
    archivePrefix = "arXiv",
    primaryClass = "astro-ph.IM"
}

@article{shorten_surveyimagedataaugmentation_2019,
    author = "Shorten, Connor and Khoshgoftaar, Taghi M.",
    title = "{A Survey on Image Data Augmentation for Deep Learning}",
    journal = "J. Big Data",
    volume = "6",
    pages = "60",
    year = "2019",
    doi = "10.1186/s40537-019-0197-0"
}

@book{maggiore_gravitational_2008,
    author = "Maggiore, Michele",
    title = "{Gravitational Waves: Theory and Experiments}",
    publisher = "Oxford University Press",
    year = "2007",
    doi = "10.1093/acprof:oso/9780198570745.001.0001"
}

@article{mehta_accurate_2017,
    author = "Mehta, Ajit Kumar and others",
    title = "{Accurate inspiral-merger-ringdown gravitational waveforms for nonspinning black-hole binaries including the effect of subdominant modes}",
    journal = "Phys. Rev. D",
    volume = "96",
    pages = "124010",
    year = "2017",
    doi = "10.1103/PhysRevD.96.124010",
    eprint = "1708.03501",
    archivePrefix = "arXiv",
    primaryClass = "gr-qc"
}

@article{mehta_testsgeneralrelativitygravitationalwaveobservationsusingflexibletheoryindependentmethod_2023,
    author = "Mehta, Ajit Kumar and others",
    title = "{Tests of general relativity with gravitational-wave observations using a flexible theory-independent method}",
    journal = "Phys. Rev. D",
    volume = "107",
    pages = "044020",
    year = "2023",
    doi = "10.1103/PhysRevD.107.044020",
    eprint = "2203.13937",
    archivePrefix = "arXiv",
    primaryClass = "gr-qc"
}

@article{khan_frequencydomaingravitationalwavesnonprecessingblackholebinariesiiphenomenologicalmodeladvanceddetectorera_2016,
    author = "Khan, Sebastian and others",
    title = "{Frequency-domain gravitational waves from nonprecessing black-hole binaries. II. A phenomenological model for the advanced detector era}",
    journal = "Phys. Rev. D",
    volume = "93",
    pages = "044007",
    year = "2016",
    doi = "10.1103/PhysRevD.93.044007",
    eprint = "1508.07253",
    archivePrefix = "arXiv",
    primaryClass = "gr-qc"
}

@article{garcia-quiros_multimodefrequencydomainmodelgravitationalwavesignalnonprecessingblackholebinaries_2020,
    author = "Garc\'ia-Quir\'os, Cecilio and others",
    title = "{Multimode frequency-domain model for the gravitational wave signal from nonprecessing black-hole binaries}",
    journal = "Phys. Rev. D",
    volume = "102",
    pages = "064002",
    year = "2020",
    doi = "10.1103/PhysRevD.102.064002",
    eprint = "2001.10914",
    archivePrefix = "arXiv",
    primaryClass = "gr-qc"
}

@article{pan_inspiral-merger-ringdown_2011,
    author = "Pan, Yi and others",
    title = "{Inspiral-merger-ringdown waveforms of spinning, precessing black-hole binaries in the effective-one-body formalism}",
    journal = "Phys. Rev. D",
    volume = "89",
    pages = "084006",
    year = "2014",
    doi = "10.1103/PhysRevD.89.084006",
    eprint = "1307.6232",
    archivePrefix = "arXiv",
    primaryClass = "gr-qc"
}

@article{broeck_phenomenologyamplitudecorrectedpostnewtoniangravitationalwaveformscompactbinaryinspiralsignaltonoiseratios_2007,
    author = "Van Den Broeck, Chris and Sengupta, Anand S.",
    title = "{Phenomenology of amplitude-corrected post-Newtonian gravitational waveforms for compact binary inspiral: Signal-to-noise ratios}",
    journal = "Class. Quant. Grav.",
    volume = "24",
    pages = "155--176",
    year = "2007",
    doi = "10.1088/0264-9381/24/1/009",
    eprint = "gr-qc/0607092",
    archivePrefix = "arXiv",
    primaryClass = "gr-qc"
}

@article{heisenberg_simultaneouslysolving08tensionslatedarkenergy_2023,
    author = "Heisenberg, Lavinia and Villarrubia-Rojo, Hector and Zosso, Jann",
    title = "{Simultaneously solving the $H_0$ and $\sigma_8$ tensions with late dark energy}",
    journal = "Phys. Dark Univ.",
    volume = "39",
    pages = "101163",
    year = "2023",
    doi = "10.1016/j.dark.2022.101163",
    eprint = "2201.11623",
    archivePrefix = "arXiv",
    primaryClass = "astro-ph.CO"
}

@article{heisenberg_canlatetimeextensionssolve08tensions_2022,
    author = "Heisenberg, Lavinia and Villarrubia-Rojo, Hector and Zosso, Jann",
    title = "{Can late-time extensions solve the $H_0$ and $\sigma_8$ tensions?}",
    journal = "Phys. Rev. D",
    volume = "106",
    pages = "043503",
    year = "2022",
    doi = "10.1103/PhysRevD.106.043503",
    eprint = "2202.01202",
    archivePrefix = "arXiv",
    primaryClass = "astro-ph.CO"
}

@software{alexnitz_gwastropycbcv233releasepycbc_2024,
    author = "Nitz, Alexander H. and others",
    title = "{gwastro/pycbc: v2.3.3 release of PyCBC}",
    year = "2024",
    doi = "10.5281/zenodo.10473621"
}

@article{moore_gravitationalwavesensitivitycurves_2015,
    author = "Moore, Christopher J. and Cole, Robert H. and Berry, Christopher P. L.",
    title = "{Gravitational-wave sensitivity curves}",
    journal = "Class. Quant. Grav.",
    volume = "32",
    pages = "015014",
    year = "2015",
    doi = "10.1088/0264-9381/32/1/015014",
    eprint = "1408.0740",
    archivePrefix = "arXiv",
    primaryClass = "gr-qc"
}

@article{rham_massivegravity_2014,
    author = "de Rham, Claudia",
    title = "{Massive Gravity}",
    journal = "Living Rev. Rel.",
    volume = "17",
    pages = "7",
    year = "2014",
    doi = "10.12942/lrr-2014-7",
    eprint = "1401.4173",
    archivePrefix = "arXiv",
    primaryClass = "hep-th"
}

@article{hinterbichler_theoreticalaspectsmassivegravity_2012,
    author = "Hinterbichler, Kurt",
    title = "{Theoretical Aspects of Massive Gravity}",
    journal = "Rev. Mod. Phys.",
    volume = "84",
    pages = "671--710",
    year = "2012",
    doi = "10.1103/RevModPhys.84.671",
    eprint = "1105.3735",
    archivePrefix = "arXiv",
    primaryClass = "hep-th"
}

@article{rham_gravitonmassbounds_2017,
    author = "de Rham, Claudia and Deskins, J. Tate and Tolley, Andrew J. and Zhou, Shuang-Yong",
    title = "{Graviton Mass Bounds}",
    journal = "Rev. Mod. Phys.",
    volume = "89",
    pages = "025004",
    year = "2017",
    doi = "10.1103/RevModPhys.89.025004",
    eprint = "1606.08462",
    archivePrefix = "arXiv",
    primaryClass = "hep-th"
}

@article{canizares_testingchernsimonsmodifiedgravitygravitationalwavedetectionsextrememassratiobinaries_2012,
    author = "Canizares, Priscilla and Gair, Jonathan R. and Sopuerta, Carlos F.",
    title = "{Testing Chern-Simons Modified Gravity with Gravitational-Wave Detections of Extreme-Mass-Ratio Binaries}",
    journal = "Phys. Rev. D",
    volume = "86",
    pages = "044010",
    year = "2012",
    doi = "10.1103/PhysRevD.86.044010",
    eprint = "1205.1253",
    archivePrefix = "arXiv",
    primaryClass = "gr-qc"
}

@book{devroye_probabilistic_1996,
    author = "Devroye, Luc and Gy{\"o}rfi, L{\'a}szl{\'o} and Lugosi, G{\'a}bor",
    title = "{A Probabilistic Theory of Pattern Recognition}",
    publisher = "Springer",
    series = "Stochastic Modelling and Applied Probability",
    volume = "31",
    year = "1996",
    doi = "10.1007/978-1-4612-0711-5"
}

@book{bishop_pattern_2006,
    author = "Bishop, Christopher M.",
    title = "{Pattern Recognition and Machine Learning}",
    publisher = "Springer",
    year = "2006",
    isbn = "978-0-387-31073-2"
}

@article{abbott_testsgr_gw150914_2016,
    author = "{Abbott, B. P. and others}",
    collaboration = "LIGO Scientific, Virgo",
    title = "{Tests of General Relativity with GW150914}",
    journal = "Phys. Rev. Lett.",
    volume = "116",
    pages = "221101",
    year = "2016",
    doi = "10.1103/PhysRevLett.116.221101",
    eprint = "1602.03841",
    archivePrefix = "arXiv",
    primaryClass = "gr-qc"
}

@article{aasi_advancedligo_2015,
    author = "{Aasi, J. and others}",
    collaboration = "LIGO Scientific",
    title = "{Advanced LIGO}",
    journal = "Class. Quant. Grav.",
    volume = "32",
    pages = "074001",
    year = "2015",
    doi = "10.1088/0264-9381/32/7/074001",
    eprint = "1411.4547",
    archivePrefix = "arXiv",
    primaryClass = "gr-qc"
}

@article{jackiw_chernsimonsmodification_2003,
    author = "Jackiw, R. and Pi, S.-Y.",
    title = "{Chern-Simons modification of general relativity}",
    journal = "Phys. Rev. D",
    volume = "68",
    pages = "104012",
    year = "2003",
    doi = "10.1103/PhysRevD.68.104012",
    eprint = "gr-qc/0308071",
    archivePrefix = "arXiv",
    primaryClass = "gr-qc"
}

@article{alexander_chernsimonsreview_2009,
    author = "Alexander, Stephon and Yunes, Nicol{\'a}s",
    title = "{Chern-Simons Modified General Relativity}",
    journal = "Phys. Rept.",
    volume = "480",
    pages = "1--55",
    year = "2009",
    doi = "10.1016/j.physrep.2009.07.002",
    eprint = "0907.2562",
    archivePrefix = "arXiv",
    primaryClass = "hep-th"
}

@article{derham_resummation_2011,
    author = "de Rham, Claudia and Gabadadze, Gregory and Tolley, Andrew J.",
    title = "{Resummation of Massive Gravity}",
    journal = "Phys. Rev. Lett.",
    volume = "106",
    pages = "231101",
    year = "2011",
    doi = "10.1103/PhysRevLett.106.231101",
    eprint = "1011.1232",
    archivePrefix = "arXiv",
    primaryClass = "hep-th"
}

@article{will_gravitonmass_1998,
    author = "Will, Clifford M.",
    title = "{Bounding the mass of the graviton using gravitational-wave observations of inspiralling compact binaries}",
    journal = "Phys. Rev. D",
    volume = "57",
    pages = "2061--2068",
    year = "1998",
    doi = "10.1103/PhysRevD.57.2061",
    eprint = "gr-qc/9709011",
    archivePrefix = "arXiv",
    primaryClass = "gr-qc"
}

@inproceedings{szegedy_rethinking_2016,
    author = "Szegedy, Christian and Vanhoucke, Vincent and Ioffe, Sergey and Shlens, Jon and Wojna, Zbigniew",
    title = "{Rethinking the Inception Architecture for Computer Vision}",
    booktitle = "Proceedings of the IEEE Conference on Computer Vision and Pattern Recognition (CVPR)",
    pages = "2818--2826",
    year = "2016",
    doi = "10.1109/CVPR.2016.308"
}

@article{kingma_adam_2015,
    author = "Kingma, Diederik P. and Ba, Jimmy",
    title = "{Adam: A Method for Stochastic Optimization}",
    journal = "Proceedings of the 3rd International Conference on Learning Representations (ICLR)",
    year = "2015",
    eprint = "1412.6980",
    archivePrefix = "arXiv",
    primaryClass = "cs.LG"
}

@article{hawking_blackholesbransdicke_1972,
    author = "Hawking, S. W.",
    title = "{Black Holes in the Brans-Dicke Theory of Gravitation}",
    journal = "Commun. Math. Phys.",
    volume = "25",
    pages = "167--171",
    year = "1972",
    doi = "10.1007/BF01877518"
}

@article{sotiriou_blackholesscalartensorgravity_2012,
    author = "Sotiriou, Thomas P. and Faraoni, Valerio",
    title = "{Black Holes in Scalar-Tensor Gravity}",
    journal = "Phys. Rev. Lett.",
    volume = "108",
    pages = "081103",
    year = "2012",
    doi = "10.1103/PhysRevLett.108.081103",
    eprint = "1109.6324",
    archivePrefix = "arXiv",
    primaryClass = "gr-qc"
}

@article{brans_machsprinciple_1961,
    author = "Brans, C. and Dicke, R. H.",
    title = "{Mach's Principle and a Relativistic Theory of Gravitation}",
    journal = "Phys. Rev.",
    volume = "124",
    pages = "925--935",
    year = "1961",
    doi = "10.1103/PhysRev.124.925"
}

@article{Heisenberg:2018vsk,
    author = "Heisenberg, Lavinia",
    title = "{A systematic approach to generalisations of General Relativity and their cosmological implications}",
    eprint = "1807.01725",
    archivePrefix = "arXiv",
    primaryClass = "gr-qc",
    doi = "10.1016/j.physrep.2018.11.006",
    journal = "Phys. Rept.",
    volume = "796",
    pages = "1--113",
    year = "2019"
}

@article{Piarulli:2025rvr,
    author = {Piarulli, Manuel and Marsat, Sylvain and S{\"a}nger, Elise M. and Buonanno, Alessandra and Steinhoff, Jan and Tamanini, Nicola},
    title = "{Parametrized test of general relativity for LISA massive black hole binary inspirals}",
    eprint = "2510.06330",
    archivePrefix = "arXiv",
    primaryClass = "gr-qc",
    doi = "10.1103/59zd-qvbd",
    journal = "Phys. Rev. D",
    volume = "112",
    number = "12",
    pages = "124044",
    year = "2025"
}

@article{LIGOScientific:2026fcf,
    author = "Abac, A. G. and others",
    collaboration = "LIGO Scientific, VIRGO, KAGRA",
    title = "{GWTC-4.0: Tests of General Relativity. II. Parameterized Tests}",
    eprint = "2603.19020",
    archivePrefix = "arXiv",
    primaryClass = "gr-qc",
    reportNumber = "LIGO-P2500066",
    month = "3",
    year = "2026"
}

@article{LIGOScientific:2026qni,
    author = "Abac, A. G. and others",
    collaboration = "LIGO Scientific, VIRGO, KAGRA",
    title = "{GWTC-4.0: Tests of General Relativity. I. Overview and General Tests}",
    eprint = "2603.19019",
    archivePrefix = "arXiv",
    primaryClass = "gr-qc",
    reportNumber = "LIGO-P2500065",
    month = "3",
    year = "2026"
}

@article{Pompili:2025cdc,
    author = "Pompili, Lorenzo and Maggio, Elisa and Silva, Hector O. and Buonanno, Alessandra",
    title = "{Parametrized spin-precessing inspiral-merger-ringdown waveform model for tests of general relativity}",
    eprint = "2504.10130",
    archivePrefix = "arXiv",
    primaryClass = "gr-qc",
    doi = "10.1103/ng8w-98sz",
    journal = "Phys. Rev. D",
    volume = "111",
    number = "12",
    pages = "124040",
    year = "2025"
}

@article{Maggio:2022hre,
    author = "Maggio, Elisa and Silva, Hector O. and Buonanno, Alessandra and Ghosh, Abhirup",
    title = "{Tests of general relativity in the nonlinear regime: A parametrized plunge-merger-ringdown gravitational waveform model}",
    eprint = "2212.09655",
    archivePrefix = "arXiv",
    primaryClass = "gr-qc",
    reportNumber = "LIGO-P2200377",
    doi = "10.1103/PhysRevD.108.024043",
    journal = "Phys. Rev. D",
    volume = "108",
    number = "2",
    pages = "024043",
    year = "2023"
}

\end{document}